%% file: main.tex
\documentclass[conference]{IEEEtran}
\IEEEoverridecommandlockouts
\usepackage{cite}
\usepackage{amsmath,amssymb,amsfonts}
\usepackage{algorithmic}
\usepackage{graphicx}
\usepackage{textcomp}
\usepackage{amsmath}
\usepackage{enumerate}
\usepackage[english]{babel}
\usepackage{xcolor}
\usepackage{filecontents}
\usepackage{makecell}
\usepackage[inline]{enumitem}
\usepackage{wrapfig}
\usepackage[most]{tcolorbox}
\usepackage{xspace}
\usepackage{subcaption}
\usepackage{graphicx} 
\usepackage{multirow}
\usepackage{booktabs}
\usepackage{array}
\usepackage{datetime}

\usepackage{tikz}
\usepackage{amsmath}
\usepackage{color, colortbl}
\usepackage{xcolor}
\usepackage{multirow}
\usepackage{url}

\usepackage{dblfloatfix}

\def\BibTeX{{\rm B\kern-.05em{\sc i\kern-.025em b}\kern-.08em
    T\kern-.1667em\lower.7ex\hbox{E}\kern-.125emX}}
\begin{document}
\title{The Toxicity Phenomenon Across Social Media}


\author{\IEEEauthorblockN{Rhett Hanscom, Tamara Silbergleit Lehman,  Qin Lv, Shivakant Mishra}
\IEEEauthorblockA{\textit{University of Colorado Boulder } \\
\{rhett.hanscom, tamara.lehman, qin.lv, mishras\}@colorado.edu}
\and
}

\maketitle

\begin{abstract}
Social media platforms have evolved rapidly in modernity without strong regulation. One clear obstacle faced by current users is that of toxicity. Toxicity on social media manifests through a number of forms, including harassment, negativity, misinformation or other means of divisiveness. In this paper, we characterize literature surrounding toxicity, formalize a definition of toxicity, propose a novel cycle of internet extremism, list current approaches to toxicity detection, outline future directions to minimize toxicity in future social media endeavors, and identify current gaps in research space. We present a novel perspective of the negative impacts of social media platforms and fill a gap in literature to help improve the future of social media platforms. 
\end{abstract}


\input{sections/02_introduction}

\input{sections/03_motivation}

\input{sections/04a_methodology}

\input{sections/04_defining_toxicity}
\input{sections/05_misinformation}

\input{sections/06_current_approaches}
\input{sections/07_demographics}
\input{sections/08_mitigation_techniques}
\input{sections/09_gaps_and_future}
\input{sections/10_conclusion}

\clearpage

\input{main.bbl}


\end{document}

%% file: sections/02_introduction.tex
\section{Introduction}
\label{sec:introduction}

Social media is no longer a simple online tool which one might use to connect to their peers, but rather has become a feature of an online world---where communication standards and practices have evolved both alongside and outside of those in more traditional interaction structures. We often see groups of people convened in an anonymous chat room behaving differently than if they were instead brought together to interact in a physical space~\cite{barlett2015anonymously,zimmerman2016online}. This illustrates the role of accountability which lessens as anonymity increases. 

Not only are a different set of rules applicable to online socialization, they appear to be rapidly changing. As virtual communities are one of humanity's more recent acquisitions, it is logical that these spaces will be tumultuous until longstanding practices and expectations can be established. This could be further complicated by the plethora of available spaces---spaces in which people can interact online  having their own individual moderation practices and community standards.
 
As a result, toxicity and negative communication patterns are taking hold across social media platforms (SMPs), occurring alongside other known crises of the virtual age such as misinformation campaigns, online culture of bots and trolls, biases towards polarizing and extreme content~\cite{nelimarkka2018social}, and the growing user distrust of public highly-visible influencer accounts~\cite{belanger2021comparison}. In order to understand the combined effects of these phenomenon it is important to first understand them individually.

Social media is ubiquitous and here to stay. However, it has been shown that social media platforms are affecting individuals in ways that have been proven to be detrimental to their health and that of their communities at large~\cite{vogels2021state, chu2024characterizing}. While virtual exchanges have become an entrenched component of human interactions, virtual spaces are not physical spaces, and it is clear that people use them differently. Recent surveys have found this difference supported by Americans, where an overwhelming 91\% listed cyberbullying and online harassment as concerning problems~\cite{vogels2021state}. 
While it is clear that online harassment is a problem, the ramifications and causes of it are not well understood.  In this paper, we provide an in-depth investigation on the factors that often contribute to negative interactions on online social media platforms with the goal of improving their designs and creating more positive online communities. The paper makes the following important contributions:

\begin{itemize}
  \item We provide a definition of online toxicity.
  \item We produce an in-depth investigation of online toxicity.
  \item We propose a novel cycle of internet extremism involving toxicity, misinformation and polarization.
  \item We systematize influential research which intersects toxicity on social media
\end{itemize}

We start by introducing motivating factors in Section~\ref{sec:overview}, defining the parameters our systematization in Section~\ref{sec:methodology}, and formally define toxicity in Section~\ref{sec:def_tox}. We then discuss the more heavily researched counterparts of toxicity, misinformation and polarization, and discuss the roles of each aspect in a cycle of internet extremism in Section~\ref{sec:misinformation}.
Current detection techniques are explored in Section~\ref{sec:current_approaches}. 
We discuss the less understood factors and details such as Demographics in Section~\ref{sec:demographics}, the impact of mitigation techniques in Section~\ref{sec:mitigation_techniques} and we finally identify important research gaps that need to be addressed in future work in Section~\ref{sec:gaps_and_future}.

%% file: sections/03_motivation.tex
\section{Motivation}
\label{sec:overview}

Discussions surrounding toxicity on SMPs have occurred ad nauseam in the public sphere, even as an exact definition on toxicity often eludes the conversation. Research spaces have likewise seen a large influx in work dedicated to toxicity classification. However, a standard definition or understanding of toxicity remains difficult to ascertain. This lack of definition often complicates moderation techniques that rely on statically defined concepts to moderate~\cite{jhaver2023personalizing}. Jhaver \emph{et al.} show how users are often confused about self moderation techniques that require a mutual understanding of the topics in question to moderate. This lack of an agreed definition for toxicity, leaves each researcher to set out to define toxicity within the scope of their own work--- making direct moderation techniques (both self and automatic) and comparisons between toxicity classifiers difficult. 

Even as research surrounding toxicity classification has progressed rapidly, less work has been published investigating the nuances of toxicity. For example, toxicity perception can vary depending on situational context and background of the one experiencing the event~\cite{sap2021annotators}. Furthermore, little work has been done to identify demographics of those receiving and producing toxic content. While the prevalence of toxicity and harassment has been shown to affect all ages and demographics in the United State s--- who exactly is creating all this toxicity is left uncertain. This gap in the literature is concerning given the frequency with which online users are directly impacted by negative online interactions~\cite{vogels2021state}. Even more so, as several studies looking at human interactions on online social media platforms identifies that toxicity does not discourage participation, and instead extends the length of the interaction~\cite{avalle2024persistent, etta2024topology}.

Toxicity often occurs alongside misinformation~\cite{cinelli2021dynamics}, polarization~\cite{vasconcellos2023analyzing} and filter bubbles~\cite{shcherbakova2022social}, further underscoring the need for a bird's eye vantage of these interrelated trends. Prior work advocated for the inclusion of the context into machine learning models given the fact that context is crucial to identify toxic content~\cite{sheth2022defining}. Our work complements this prior work by Sheth \emph{et.al.} by providing a formal definition of toxicity and delving deeper into the context surrounding toxicity to identify the root of the problem. The need for a consolidation of knowledge is echoed by the large number of fields across research and industry which are related to toxicity (natural language processing, software engineering, psychology, sociology, etc.), which would benefit from a single work defining online toxicity and synthesizing the most pertinent information relating to SMP toxicity.

%% file: sections/04a_methodology.tex
\section{Methodology}
\label{sec:methodology}

\begin{table*}[t]
\centering
\begin{tabular}{|
>{\columncolor[HTML]{C0C0C0}}c |c|}
\hline

\textbf{Defining Toxicity}  & \cellcolor[HTML]{FFFFFF} \makecell{\cite{salminen2020developing}, \cite{kim2021distorting}, \cite{fan2021social}, \\ \cite{saveski2021structure}, \cite{sheth2022defining}, \cite{almerekhi2020these}}

\\ \hline
\textbf{\begin{tabular}[c]{@{}c@{}}Toxicity Detection\\ and Classification\end{tabular}}      & 

\makecell{\cite{salminen2020developing}, \cite{muneer2020comparative}, \cite{rupapara2021impact},  \\ \cite{abro2020automatic}, \cite{kumar2021designing}, \cite{fan2021social},  \\ \cite{sheth2022defining}, \cite{akuma2022comparing} }

\\ \hline
\textbf{Toxicity and Filter Bubbles}                                                          & \makecell{\cite{nguyen2014exploring}, \cite{nechushtai2019kind}, \cite{dahlgren2021critical}, \\ \cite{gao2023cirs}, \cite{aridor2020deconstructing}, \cite{wang2022user}}

\\ \hline
\textbf{\begin{tabular}[c]{@{}c@{}}Toxicity, Polarization\\  and Misinformation\end{tabular}} & \makecell{\cite{au2022role}, \cite{pascual2021toxicity}, \cite{cinelli2021dynamics}, \\ \cite{salminen2020topic}, \cite{simchon2022troll}}

\\ \hline
\textbf{Toxicity Demographics}  & \makecell{\cite{vogels2021state}, \cite{cosma2020bullying}, \cite{ding2020profiles}, \\ \cite{pichel2021bullying}    }

\\ \hline
\end{tabular}
\caption{The most formative and cited work across several research areas which intersect with toxicity on SMPs. For each topic matter.}
\label{tab:most_cited}
\end{table*}

`Toxicity' as a term has fully entered the research vernacular.
Table~\ref{tab:most_cited} lists the most cited papers across a number of broad issues related to toxicity on SMPs. 
These are all the piece of work we considered in our work. The rationale for how we selected these pieces of work is detailed next. 

Due to the volume of research involving both toxicity on social media as well of the large number of research areas which involve online toxicity, not every piece could be considered in the scope of this characterization. Instead, only work published at scientific venues which are known to explore social media, psychology, and computer science through rigorous peer-review processes were included within this work---except where otherwise noted. Papers with 40 or more citations and published within the last five years were given preferential treatment when selecting papers for inclusion. Much of the cited work comes from IEEE/ACM Conferences, such as CSCW, ASONAM, HCI, Human Factors in Computing Systems, etc., and other well respected research organizations: Pew Research and Elsevier.

%% file: sections/04_defining_toxicity.tex
\begin{figure}[h]
\vspace{-0.1in}
\begin{tcolorbox}[colback=yellow!5!white,colframe=yellow!50!black]
  \textit{Toxicity}:
interactions directed at an entity designed to be inflammatory and purposefully breed counterproductive dissension.
\end{tcolorbox}
\vspace{-0.2in}
\label{fig:toxicity_def}
\end{figure}

\section{Defining Toxicity}
\label{sec:def_tox}

\begin{figure}[b]
        \begin{center}
    \includegraphics[width=0.35\textwidth]{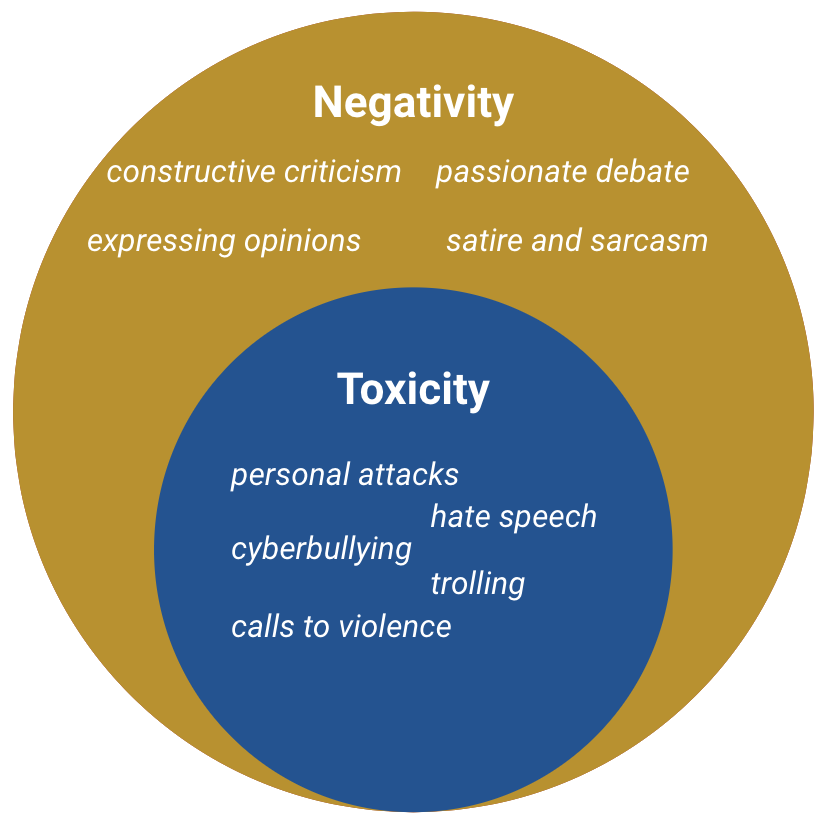}
  \caption{Difference between toxicity and negativity.}
  \label{fig:venn}
  \end{center}
\end{figure}

We propose to define toxicity as any interaction directed at an entity designed to be inflammatory and purposefully breed counterproductive dissension.  
Toxicity can be directed at a community, ideal, organization or another entity, and need not be directed at a specific individual. Based on the proposed definition, toxicity always places itself in opposition to another entity, and therefore toxicity must always have a specific target, differentiating toxicity and negativity. The nuances of toxicity and negativity are illustrated in Figure~\ref{fig:venn}.

According to the proposed definition, the following terms fall under the umbrella of toxic behavior: hate speech, cyberbullying, threats, vulgarity, and highly-polarized content, so long as these actions are taken in opposition to an entity.
Non-toxic negativity can include constructive criticism, the expression of dissenting opinions, reporting misinformation, satire and sarcasm, passionate debate, and public call-outs. Negativity can be harsh but is often the result of a desire for positive change, a feature that is not present in toxic behavior.

Toxicity does not exist free of context and is subject to be perceived at different levels by those who experience it~\cite{fiske2018controlling, croom2011slurs}. The proposed definition brings in a measure of objectivity to the highly subjective nature of toxicity. Purpose is one key factor in identifying toxicity, which sets it apart from general discourse and negativity. Negativity exists separately from toxicity and may be part of healthy online interaction. While all toxicity is negative, not all negativity is toxic. In this paper, we use the proposed definition of toxicity to construct a collection of prior work with the goal of identifying the root sources of the problem.

\begin{table*}
\centering
\begin{tabular}{|l|r|r|l|}
\hline
\rowcolor[HTML]{C0C0C0} 
          & \multicolumn{1}{l|}{\cellcolor[HTML]{C0C0C0}\begin{tabular}[c]{@{}l@{}}Mention `Toxicity'\\ and `Social Media'\end{tabular}} & \multicolumn{1}{l|}{\cellcolor[HTML]{C0C0C0}\begin{tabular}[c]{@{}l@{}}`Define Toxicity'\\ and `Social Media'\end{tabular}} & \begin{tabular}[c]{@{}l@{}}Offer Definition \\ of Toxicity\end{tabular} \\ \hline
2023-2024 & 21,400                                                                                                                     & 51                                                                                                                          & \multicolumn{1}{r|}{8}                                                  \\ \hline
2018-2024 & 82,600                                                                                                                       & 102                                                                                                                         & \cellcolor[HTML]{C0C0C0}                                                \\ \hline
All Time  & 899,000                                                                                                                    & 125                                                                                                                          & \cellcolor[HTML]{C0C0C0}                                                \\ \hline
\end{tabular}
\caption{Google Scholar search results for literature on toxicity.}
\label{tab:toxicity_survey}
\end{table*}
 
\subsection{Previous Attempts to Define Toxicity}
\label{prev_attempts}

Table~\ref{tab:toxicity_survey} illustrates the current state of toxicity in research. A Google Scholar search for articles mentioning both `toxicity' and `social media' across all time returns some 900,000 items. However, if we narrow the search down to `social media' and at least one of `defining toxicity' and `define toxicity' only 125 articles are returned with 51 of these being published within one year. Reviewing these recent 51 articles, only actually put forth definitions of toxicity, with most favoring either overly general definitions or highly specific definitions only meant to be contained within the bounds of a single study~\cite{recuero2024platformization, Madhyastha_Founta_Specia_2023}. 

In prior work where toxicity was defined, definitions have included: harassment, threats, obscenity, insults, identity-based hate as well as socially disruptive persuasion~\cite{sheth2022defining, salminen2020developing}. Toxicity is often used interchangeably with cyber-bullying \cite{fan2021social}. Still other work has described toxicity as behavior which demonstrates \textit{disrespect for others} \cite{kim2021distorting}. Differences in the perception of toxicity mean that experienced hard and intent to hard do not always align~\cite{almerekhi2020these}.

The Perspective API's definition of toxicity: ``a rude, disrespectful, or unreasonable comment" has been used in research to define toxicity \cite{saveski2021structure}. Some researchers have combined misinformation, polarization and toxicity into a single phenomenon \cite{sheth2022defining}. The conflation of these terms may contradict existing and supported definitions of these phenomena~\cite{vraga2020defining,kubin2021role}. In this work, we choose to build on existing research which establishes misinformation and polarization as distinct entities, and instead contextualize and position these phenomena alongside filter bubbles and toxicity as distinct components of a cycle of online extremism. This distinction is further expanded upon in Section~\ref{sec:misinformation}.

%% file: sections/05_misinformation.tex
\section{Co-Occurring Phenomena}
\label{sec:misinformation}

Online SMPs have introduced many new phenomena that have less understood consequences on individuals and as a result thrust a large variety of studies into their impact and interactions. Among several of the phenomena studied are bots and trolls, misinformation and polarization, and the filter bubble effect \cite{nguyen2014exploring}. We describe the current state of the art for these phenomena and propose a novel perspective on the interaction between misinformation, polarization and toxicity which we call \emph{the cycle of internet extremism}.

\subsection{Bots and Trolls}
\label{sec:botstrolls}
To understand how people are affected by toxicity, we must also understand bots and trolls, key players in online SMP spaces. 
A \textbf{troll account} has the primary objective of spreading discourse, mostly commenting incendiary content with the intent to cause the most extreme reaction. \textbf{Bot accounts} are run and managed by sophisticated software, capable and ready to capitalize on nuanced conversations in an attempt to sway public opinion. These accounts are often members of a larger \textbf{bot swarm}, which operates towards a unified goal. Currently, even though bot accounts add up to only 5\% of all Twitter profiles, it is estimated that 20-30\% of Twitter content consumed in the United States is created by bots~\cite{varanasi}. This percent disparity demonstrates how great the reach of these bot accounts is. 

Bot creators have weaponized this opportunity in a number of ways, including: attempts to sway election results~\cite{dutta2021analyzing}, rehabilitate brand reputation~\cite{aydemir2020social}, and boost album sales~\cite{boyer_2021}. This evidence supports the idea that bots may not necessarily contribute to toxic behavior, 
while a troll account is specifically designed to lead to it.

\subsection{Filter Bubbles, Misinformation and Polarization}
\label{sec:misin_polar}

An integral component of understanding the state of SMPs is the spike in misinformation which has flooded the Internet in recent years~\cite{allcott2019trends}. We have seen the Russian government involved in a number of instances of politically motivated misinformation campaigns in which they purposely spread false information and amplify highly-polarized opinions about upcoming elections~\cite{alba_frenkel_2019, washingtonpost, woolley2017computational, dutta2021analyzing}. While reports vary on the effectiveness of these campaigns, they are a clear and a looming threat to the integrity of online discourse. 

The spread of misinformation has become increasingly complex as most SMPs have introduced features meant to encourage users to consume news on their systems. Facebook and Twitter have become enormous providers of news, even without producing content of their own or implementing well established editing practices---compared with news organization counterparts which abide by their ethical journalistic responsibilities. Research shows that accounts posting misinformation are significantly more likely to be bots and are able to quickly saturate a virtual space with content~\cite{shao2018spread}.

The influx of misinformation on SMPs has been widely documented yet continues to spread rapidly.
It has been shown that social media users place importance on vetting news articles shared on social media, placing particular significance on the news source and content~\cite{pidikiti2020understanding,pidikiti2022understanding}. However, people make decisions about news veracity quickly and often do not weigh all the information they flag as theoretically pertinent when making snap decisions about trustworthiness in actuality~\cite{klein2018people, pidikiti2020understanding,pidikiti2022understanding}. These studies show that users are more susceptible than they would anticipate, even when armed with integrity-discerning practices.

Furthermore, the use of topic-based recommendation systems can lead to the formation of filter bubbles (often referred to as `echo chambers') \cite{nguyen2014exploring, gao2023cirs, aridor2020deconstructing}. Online communities that are formed with such strict adherence to a single ideal or notion naturally lead to behavior in which non-homogeneous contributions are not accepted. Online \textbf{filter bubbles} create circles where misinformation can thrive. Individuals are often suggested media content of a single, usually extreme, perspective. Users are then led to believe that all other viewpoints are irrelevant. This effect creates a problem, as growth and healthy discussions hinge on the introduction of novel ideas and new perspectives.  As a result, filter bubbles can push communities to focus on highly polarized stances as opposed to understanding the nuances of complicated issues \cite{spohr2017fake}. Filter bubbles reinforce notions from psychology such as: people will generally seek supportive information over challenging information, people will gravitate towards like-minded individuals~\cite{dahlgren2021critical}. As AI tools are increasingly integrated into our daily lives, the use of AI in curating content through recommendation algorithms has lead to highly homogeneous media diets like those seen in filter bubbles \cite{nechushtai2019kind}.

Two types of polarization currently play critical roles in social media: ideological and affective. Ideological polarization is defined as the "divergence of ideals along party lines”~\cite{kubin2021role}, while affective polarization refers to the degree to which a user feels aligned or opposed to individuals with different political identities~\cite{kubin2021role}. Simply put, ideological polarization is the divergence in political ideals, while affective polarization is empathy for peers viewed as out-group members.

The relationship between misinformation and polarization has been widely documented in contemporary work \cite{au2022role}. The availability and propagation of misinformation has been shown to be a critical component of fostering high discourse and polarization in online communities~\cite{azzimonti2022social}. Especially for users less experienced in online spaces, exposure to misinformation can have lasting effects on how active these users will be within their communities in both the short and long term~\cite{wang2021analyzing}. These users are often galvanized to participate in public discourse, not realizing the highly politicized bias and deceptions embedded in the media they consume. In fact, some prior work has found that interacting with the accounts run by the Russia-backed election interference campaign prior to the 2016 U.S. presidential election led to Twitter users doubling down on their online behavior~\cite{dutta2021analyzing}. Many of these users were shown to spend more time on Twitter after run-ins with these Russian accounts~\cite{dutta2021analyzing}. 

An important result from prior studies established that Twitter users who came into contact with interference campaigns have also shown an increase in polarization, hinting that these individuals may have been pushed to extreme ideology by the Russian contacts~\cite{dutta2021analyzing}. 
Though the impact of the Russian misinformation campaign on the election outcome itself is still highly contentious, the fact that users' behavior changed is undeniable~\cite{eady2023exposure, dutta2021analyzing}. \cite{pascual2021toxicity} also drew a link between polarization and greater toxicity. Other work, on YouTube, found that online debate tend to gravitate towards toxicity~\cite{pascual2021toxicity} and that news topic affect user toxicity levels~\cite{salminen2020topic}.

\subsection{Cycle of Internet Extremism and Filter Bubbles}
\label{filter_tox}

\begin{figure}[t]
  \begin{center}
    \includegraphics[width=0.45\textwidth]{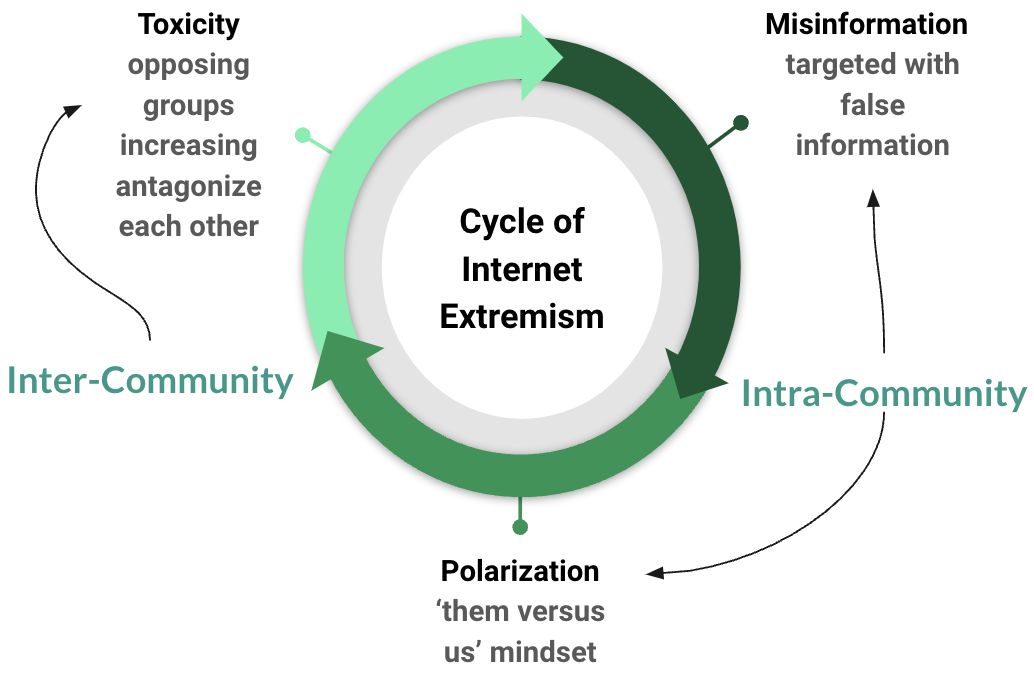}
  \end{center}
  \caption{Cycle of toxicity, misinformation, and polarization.}
  \label{fig:cycleofIE}
\end{figure}

We posit that within some communities a \emph{cycle of internet extremism} exists and it is often enhanced by filter bubbles (shown in Figure~\ref{fig:cycleofIE}). Filter bubbles, misinformation, and polarization are sometimes conflated to a single phenomenon \cite{sheth2022defining}. However, we propose to consider these topics independently and as part of a cycle of internet extremism that leads to toxicity. Within some online communities, misinformation is propagated leading to greater ideological polarization (enhanced by filter bubbles), which eventually results in increased affective polarization, as individuals are increasingly placed in an "us versus them" mindset. As polarization among community members increases, toxicity is then created as a byproduct of this cycle. This is supported by work has found that on Twitter, users post more toxic content when interacting with users with whom they do not share mutual connections \cite{saveski2021structure}. These malbehaviors (toxicity, misinformation, and polarization) continue to pressure community members towards extremes so long as they continue their participation.

Given that each phenomenon depicted in Figure~\ref{fig:cycleofIE} can occur independently outside of this cycle, prior work has studied several combinations of these. For example, the relationship between filter bubbles and misinformation has been found to influence users' gullibility~\cite{rhodes2022filter, lakshmanan2022quest, tomlein2021audit}. Filter bubbles and polarization, has also been found to be linked to "tribal mindset" in online SMPs~\cite{chitra2020analyzing, keijzer2022complex,wolfowicz2023examining}. A number of other pieces of work have also looked at misinformation, polarization and filter bubbles altogether \cite{spohr2017fake, interian2023network, simchon2022troll}, and  found a tight link among them. Though significant follow up research is required to measure the effect polarization may have on toxicity, the effect of filter bubbles on toxicity is only beginning to be explored in existing literature which found that filter bubbles can often interfere with user's critical thinking abilities~\cite{shcherbakova2022social}. 

While a lot of work has investigated the interaction of misinformation and polarization, toxicity lacks a formal definition and as a result it is more difficult to track. Based on the toxicity definition in our work, toxicity tends to appear in an inter-community manner, making its identification more difficult. Misinformation and polarization are both generally occurring at an intra-community level, through interactions with like-minded peers. In contrast, toxicity spews outwards through polarized interactions with individuals outside their in-group setting. This makes toxicity somewhat more difficult to trace, as it's decentralized and can seem to crop up everywhere, unlike misinformation and polarized content which congregate unchallenged within community bounds.
When considered in this framework, toxicity can be viewed as a direct effect of greatly affective polarization, a consequence of diminished empathy for those holding contrary ideals. This view solidifies each misbehavior into a distinct action, though they are highly convoluted in reality.

%% file: sections/06_current_approaches.tex
\section{Current Approaches to Toxicity Detection}
\label{sec:current_approaches}

Machine learning (ML) models are typically used for toxicity detection on SMPs. Prior work has offered robust evaluations of which combination of ML models and feature extraction techniques lead to the most accurate methods~\cite{abro2020automatic, muneer2020comparative}. 

Modern ML approaches to classification cannot work with text alone. Text must be converted into numerical data in order to be classified by most ML models. This is a critical step, as the information which is generated in order to train a classifier will directly shape the classifiers behavior. This step uses a feature generation tool to make the conversion. The best feature generation tool depends on both the data and the type of inferences made~\cite{abro2020automatic, muneer2020comparative}.  

Toxicity classification has traditionally been considered as supervised learning task -- meaning that models are trained with pre-labeled examples of toxicity, as tasked with learning from these examples in order to later classify documents. \cite{abro2020automatic} as well as \cite{muneer2020comparative} each tested a number of popular feature generation and supervised ML model combinations on pools of pre-labeled tweets. \cite{abro2020automatic} paired eight supervised ML modes and three feature generation tools to each separately classify text documents into 'hate speech', 'not offensive', and 'offensive but not hate speech'.  One key takeaway from this work is that no one combination of supervised ML model and feature generation tool stands as superior to the rest in all ways, but rather certain combinations are best at specific tasks and datasets. The community-level turning of toxicity classifiers, which affects minority individuals differently, can improve the effectiveness of generally trained models (such as Jigsaw's Perspective API)~\cite{kumar2021designing}. Sheth \emph{et al.}'s 2022 underscored the critical role of context in toxicity classification \cite{sheth2022defining}. 

One widely used ML model is Logistic Regression. Logistic Regression constructs a separate hyper-plane between two datasets utilizing the logistic function \cite{patchin2011traditional}. Logistic Regression takes a number of weighted features as input and produces a binary decision on classification tasks. Another high performing model, Random Forest, utilizes a number of decision trees and randomized voting mechanisms, averaging outcomes to enhance predictive accuracy and reduce over-fitting \cite{dadvar2013improving}. Support Vector Machine (SVM) is another supervised machine learning classifier often used for text classification tasks. SVM transforms an initial feature set by projecting it into a custom-defined kernel-generated space with higher dimensions. Its objective is to maximize the margin, or distance, between the categories by identifying support vectors. These support vectors are the samples from each category that are closest to the separating hyperplane initially approximated by SVM. In \cite{joachims1998text} SVM was shown to decrease in accuracy as data size increased. 

While the Support Vector Machine algorithm in partnership with bigram feature generation most successfully classified documents for \cite{abro2020automatic}, a similar analysis of comparable models produced by \cite{muneer2020comparative} instead found Logistic Regression most accurate when detecting cyberbullying on Twitter. Decision trees have also shown to pair favorably with classification on short documents such as tweets~\cite{akuma2022comparing}. 

Due to it's relative rarity when compared with non-toxic internet content, toxicity classifiers often struggle with imbalanced data sets. Over-sampling of toxic comments with SMOTE is one means of balancing training data \cite{rupapara2021impact}. Other solutions include K-Nearest Neighbors sampling, SMOTE variants (Borderline-SMOTE, DBSMOTE, etc.) and under-sampling~\cite{wang2021review}.

Unsupervised ML models are \textit{not} trained on labeled data but rather pre-trained on large corpuses and taught to infer the structure, context, and semantics of the text by predicting missing words in a sentence, understanding relationships between words, and generating coherent text. Most recently, unsupervised ML models -- such as GPT and BERT -- have become sufficiently sophisticated as to allow for their application to supervised learning tasks. New applications of unsupervised ML models to the supervised learning task of toxicity classification have proved promising \cite{sun2019fine}. This is accomplished through fine-tuning the pre-trained models on smaller, often labeled datasets, offering all the insight of the labeled data alongside the great deal of contextual knowledge gained during the pre-training.

\cite{rasmussen2024super} put forth a super-unsupervised hybrid model which combines features from both approaches to label data without a single pre-labeled training set. When compared with widely used Perspective API (discussed further in section~\ref{ssec:plat_strat}), \cite{rasmussen2024super} was able to accurately identify 'political hate'. Moving away from labeled dataset, could lead to more successful and less prejudiced toxicity classification as annotator bias has been shown to introduce bias into supervised ML models \cite{sap2021annotators}. Other work has found unsupervised learning models, such as GPT, to perform exceptionally well with short text document classification \cite{balkus2022improving}, a problem which often arises when classifying social media posts. BERT has similarly be shown to excel at toxicity classification tasks with a large number of complex features \cite{salminen2020developing, fan2021social}. Beyond simply detecting incoming toxicity on SMP, Almerekhi \emph{et al.} (2020) found success in predicting non-toxic comments which would incur toxic replies within specific communities on Reddit using a neural network in conjunction with BERT~\cite{almerekhi2020these}. 

%% file: sections/07_demographics.tex
\section{Demographics and Behaviors}
\label{sec:demographics}



Though online harassment can occur in a multitude of virtual spaces, a majority of individuals have reported their most recent harassment as having occurred on social media~\cite{vogels2021state}. Given the rise of social media usage among adults 65 and older~\cite{perrin2015social}, it is clear that online harassment is not a problem only for children and young adults, but rather it is a pressing concern for any individual who finds themselves interacting with others online. 


\subsection{Users Receiving Toxic Content}
\label{ssec:demographics_1}

Prior studies find that a staggering number of internet users experience online harassment on a daily basis. Multiple works from  across the globe have found that approximately 30\% of adolescence are actively being bullied \cite{ding2020profiles, pichel2021bullying}. Another work found that nearly half of young people who reported being cyber-bullied also report traditional bullying, echoing the severity of this issue \cite{cosma2020bullying}. A 2021 Pew Research Center report, which surveyed only adults, found that 41\% of Americans report experiencing online harassment, and 25\% report experiencing severe online harassment~\cite{vogels2021state}. If only young adults (aged 18 to 29) are considered, 64\% report experiencing online harassment, with 48\% of them reporting severe harassment~\cite{vogels2021state}. 
 Given the disproportionate amount of young people facing online harassment, this problem will continue to grow as younger generations are more plugged into social media at large.


In addition to age, gender and race have been found to be linked to harassment. Though men report harassment at higher rates than women, more women identify gender as being part of the harassment (about half). 
While only 12\% of adults report race as having been a contributing factor in their received harassment, it is also true that Black and Hispanic Americans (about 47\%) are more likely than White Americans (about 40\%) to experience racially-driven toxicity~\cite{vogels2021state}.

The gender difference also affects the type of harassment. Men are more likely to face name-calling and physical threats, while women report higher rates of sexual harassment~\cite{vogels2021state}. This effect is particularly strong for younger women (35 and under), in which 33\% have reported having received sexual harassment online~\cite{vogels2021state}. This same survey made note of the steep rise in self-reported sexual harassment targeted at women, which has doubled since their last wide-spread survey conducted four years prior.

\subsection{Users Posting Toxic Content}
\label{ssec:demographics_2}

Unfortunately very little emphasis has been placed on understanding who are online harrasers and what demographics they present~\cite{mall2020four, vaidya2024analysing}. Mall \emph{et al.} divide Reddit users who post toxic comments into four subcategories in increasing order of occurrence: \begin{enumerate*}
    \item \label{lab:fickle-minded} Fickle-Minded Users who often post both toxic and non-toxic comments (31.2\%),
    \item \label{lab:pacified} Pacified Users whose comments became less toxic over time (25.8\%), 
    \item \label{lab:radical} Radicalized Users whose comments became increasingly toxic over time (25.4\%), and 
     \item  \label{lab:steady} Steady Users whose toxicity levels are stable in comments over time (17.6\%).  
    \end{enumerate*}

This finding supports a hypothesis from prior work that there is likely a large overlap between users experiencing and posting toxicity in virtual spaces~\cite{garaigordobil2015cyberbullying}. Another study found that exposure to toxic comments lead individuals to later post more toxic content \cite{kim2021distorting}. Note that further work is needed to explore what might lead to this phenomenon. As stated previously, without a firm understanding of the reasons to harass others online, we cannot make any assumptions for this behavior to exist. Emerging SMPs in recent years have attempted to reduce toxic content by shrinking the divide between their user's physical and virtual lives, as evidenced by platforms such as TikTok, BeReal and Instagram. However, not much progress has been made in terms of reducing toxic content.  

One possible explanation for the overlap of users could be explained by the findings of a more recent study, which finds that only about 1.4\% of users infuse less toxicity into online communities than they receive~\cite{vaidya2024analysing}. Vaidya \emph{et al.} classify Twitter users based on their spread of toxic content using Google's Perspective API scores (see Section~\ref{ssec:tox_id} for details)~\cite{perspective_1}. In this study, the authors evaluate a period of 12 months in 2017 of a Twitter dataset and look at the spread of toxic content within the network. They find that only 1.4\% of users share less toxic content than they receive, that 5.3\% of users infuse more toxic content than what they receive and that about 93.3\% of users simply share the same amount of toxic content they receive. These results show that toxic content is often shared more frequently than its counterpart making the identification of toxic content creators even more difficult.  

In contrast,
the neighborhood-based NextDoor has been investigated to understand how xenophobia could have become such a fixture on a service which was intended to foster community among people living nearby~\cite{bloch2022aversive, kurwa2019building, payne2017welcome}. Some researchers have surmised this effect was due to the tone and culture established by NextDoor's platform features, which encouraged community self-policing)
~\cite{payne2017welcome}.


%% file: sections/08_mitigation_techniques.tex
\section{Mitigation Techniques}
\label{sec:mitigation_techniques}

As awareness and understanding of online toxicity progresses, so have mitigation tools. Besides developing systems for identifying harmful posts, researchers have investigated systems which can minimize the emergence for toxicity.

\subsection{Social Media Platform Strategies}
\label{ssec:plat_strat}

Currently most major SMPs have formal stances which set themselves in opposition to online harassment, and offer their users means of reporting alleged harassment~\cite{twitter1, twitter2, meta, reddit1}. 
Some current intervention strategies which empower users and communities to moderate themselves appear effective. 

In 2021, Twitter introduced a new feature that evaluates `replies', in which when one user responds to another user's post, it checks for potentially harmful language~\cite{diaz_2021}. If the reply is deemed harmful, Twitter prompts the user to examine their reply and consider rephrasing or refraining from commenting. Diaz \emph{et.al.} found that 34\% of users chose to delete their response at this junction~\cite{diaz_2021}. 


In 2019 Facebook enabled greater control to Facebook group moderators by allowing them to use the tool to identify toxic posts~\cite{hanbury}. 
Prior to this change, private Facebook group users were able to post harmful content. Given that these groups operated in a `black box' outside of Facebook's purview, they did not need to uphold to Facebook's harassment policies. While these strategies are steps in the right direction, there have been no studies on Facebook effectiveness at combating toxic content.


\textit{League of Legends} is a successful example on how to combat online harassment by handling peer conflict via a jury of your peers.  
In this online video game, in which teams of players are set against each other in online combat and are allowed to communicate via a chat system, players often face a `jury of their peers' to review the chat log and decide if behavior warrants further investigation. If escalated, the video game resembles the model of the jury in the United States, in which randomly selected players are selected to decide the outcome of the conflict. A study looking into the effectiveness of this process found that 
an overwhelming 92.98\% were convicted as `guilty'~\cite{ehrett2016judiciaries}. 

While existing SMPs have made some efforts in combating online toxicity, 82\% of Americans in 2021 did not believe that SMPs had done a `good job' in addressing toxicity~\cite{vogels2021state}. In fact, 55\% of those surveyed listed online harassment as a major problem (and an overwhelming 91\% agreed it was at least a minor concern). These same individuals are looking to SMPs to protect them from potential harmful content, but SMPs continue 
to prioritize content monetization over user experience and safety. Unfortunately, only a minority of Americans currently support holding SMPs legally responsible for content posted on their services. 

\subsection{Toxicity Identification}
\label{ssec:tox_id}
In recent years there have been several pieces of work leveraging more robust ML capabilities for toxicity identification~\cite{he2023you,pavlopoulos2021semeval,li2024hot,salminen2020developing}. Google's Perspective API is one these frameworks which uses machine learning to asses comment toxicity across a large number of languages~\cite{perspective_1}. 
Perspective API toxicity definition is "a rude, disrespectful, or unreasonable comment that is likely to make people leave a discussion.". The API collects a number of attributes, including identity attacks, insults, profanity and threat, to summarize them into one toxicity score. The problem with this toxicity definition is that it is rooted on explicit words, making it vulnerable to human biases and agnostic to context~\cite{goyal2022your}.

The importance of an agreed upon definition for toxicity cannot be understated. Perspective API has been used to manage the comment moderation of a number of high-profile websites such as The New York Times and Disqus~\cite{perspective_2}. However, given the fragility of the toxicity definition, prior work has shown how they can significantly lower Perspective API detection accuracy by slightly modifying words in a sentence~\cite{hosseini2017deceiving, jain2018adversarial}. Furthermore, given that the classifier is built on human annotations, it is critical to formally define toxicity to avoid annotator's biases~\cite{goyal2022your}.

\subsection{Research Strategies}
\label{ssec:research_strat}

Academia allows for the exploration of more abstract means of fighting toxicity. 
Some ways researchers are combating toxicity lie in fostering a greater understanding of its causes. Only through a firm comprehension of the contributing factors to toxicity in online spaces we can begin to combat it. Researchers have been looking at four different factors to combat toxicity: echo chamber effects, self-regulation, 
anonymity vs. accountability, and size of communities. 

\subsubsection{Echo Chamber Effects}
\label{ss8_1}
One key idea is that as recommendation systems are widely deployed in SMPs, polarization among users in different circles grows and hence toxicity rises~\cite{grossetti2020community}. 
These echo chambers are often formed by (or reinforced by) a recommendation system to introduces users to new accounts or content. 
One way to break the filter bubble-building effect of recommendation systems, is to build systems which 
 ensure that user recommendations create communities with somewhat varied ideologies (such as Community Aware Recommendations)~\cite{grossetti2020community}. 



\subsubsection{Self-Regulation in Online Communities}
\label{ssec:selfreg}


Prior work investigating self-moderating communities has found that self-moderation is more effective in smaller communities and that larger communities are more amenable to automated harassment detection tools~\cite{seering2020reconsidering}. The Twitch channel found four times as many cases of harassment flagged by bots as compared to those flagged by users, even when the same number of bots and users were present. This result might indicate that users are less likely to initiate the action of flagging harassment and highlights large platforms problems in moderating large amounts of content. 

\subsubsection{Anonymity vs Accountability Trade-Off}
\label{ssec:anon}


To better understand the trade-off between anonimity and accountability,
one study followed the news source \textit{Huffington Post} through three distinct policies on comment anonymity~\cite{moore2021deliberation}. Identity rules on \textit{Huffington Post} first allowed users to post comments without creating an account, allowing them to assign any name they wished to their comment. The next phase mandated that users create an account before commenting, allowing the users to utilize `sustained pseudonyms' when posting comments. The final policy 
removed the internal user system and instead mandated that users linked their comments to their Facebook profiles. This last phase meant that users real-life identities would now be tied to their posts. These three phases span the entire spectrum from truly anonymous to truly accountable. Moore found that as the anonymity restrictions increased, there were fewer comments posted. While there were fewer overall comments during the second phase, 
the cognitive complexity of comments in this phase sharply increased. This increase in the cognitive complexity of comments was present in the final phase as well, when Facebook profiles were used. These findings indicate that users are less encouraged to post their honest opinions when they feel they will be held accountable by their peers.

\subsubsection{Smaller Communities}
\label{ssec:smaller}
%
Mirroring physical structures, placing an emphasis on smaller communities can induce a greater sense of accountability among online peers, decreasing the propensity to inflammatory and polarizing content. 
In a classroom setting, there is a link between the ratio of students to teachers and class environment. When a class has fewer student-to-teacher ratios, positive student-teacher interactions tend to increase~\cite{folmer2010class}.
Another study found that in online classrooms, discussions held in small groups led to higher rates of student participation~\cite{yang2022investigating}. 
Although social media is biased in favor of larger more complex social networks, 
one piece of work which analyzed six different dark web forums found that despite the relatively small size of these networks and the overall anonymity of the user base, small-scale `influencer' nodes were still incredibly important in maintaining an active dialogue within the forum~\cite{pete2020social}. This finding suggests that no matter the size of the social network, there is still some number of critical nodes whose role is to bring together the less active and social members of a community, even when network-wide `influencers' are not present.

%% file: sections/09_gaps_and_future.tex
\section{Gaps and Future Work}
\label{sec:gaps_and_future}


Most prior work on SMP analysis has focused on the impact on users rather than the design components of the platform. 
SMPs are often considered a passive party, through which an individual or institution can post or engage with a community. 
The ramifications of an SMP do not carry consequences  when users are negatively impacted. As a result, the individual or community affected must develop coping mechanisms when faced with online harassment. This sentiment is present across work which focuses on the intersection of mental health and social media usage.
To instigate positive change we present several avenues for future work in which SMPs can be improved by diminishing online toxicity. 

\subsection{Perpetrator Demographics and Motivations}
\label{ssec:gaps_1}

A large gap in the literature is the study of the motivations and demographics of online aggressors. In part, one reason for this gap is the difficulty in gathering this information. However, without these critical components, it is difficult to move beyond the current state 
towards active strategies to disincentivize users from engaging in abusive communication. 


With such high rates of victimization, either a small number of SMP users devote a large portion of their online presence to promoting toxicity and spreading online harassment, or there must be considerable overlap between cyberbully victims and perpetrators. One prior research supports the notion that online harassment creates a cycle where victims progress to become perpetrators~\cite{garaigordobil2015cyberbullying}. 
However, that study showed that only 15.5\% of users who have experienced online bullying go on to similarly assail others~\cite{garaigordobil2015cyberbullying} -- and yet so little is know about toxic SMP users in the research space.

\subsection{Changing Social Media Platform Focus}
\label{ssec:gaps_3}

The focus of addressing online toxicity must be placed on social media platforms themselves. 
Toxicity, misinformation, and polarization, 
are all symptoms of unhealthy systems. With online harassment targeting millions of people, it is clear that cyberbullying is a feature of current SMPs and not just the unanticipated acts of a small minority of users~\cite{vogels2021state}. Research has found that polarized content can spread across SMPs faster than non-polarized content, highlighting one way that SMP algorithm bias can incentivize the creation of toxic content~\cite{vosoughi2018spread}. We should not introduce new features to combat them, but instead investigate what about current platforms encourages these malbehaviors to 
thrive. 
There is a need to understand what enables users to feel empowered to prey on others. 

\subsection{Feature Trade-Offs}
\label{ssec:gaps_5}
Future work must be devoted to understanding the trade-offs which shape user behavior. For instance, there is already clear evidence that people post differently on anonymous accounts than they do when they feel their user account is tied to their identity~\cite{zhang2014anonymity}. 
Another unexplored trade-off is the knowledge of online bot-based accounts. Despite the vast amount of research on bot behavior, there has not been any studies to look into toxicity content changes when interacting with bots. 
We posit that there are many more trade-offs which are yet unidentified.





\subsection{Recommendation Algorithms}

Research on SMPs recommendation systems comprise a significant amount of the work in social media analytics. Most of this work hinges on friend networks and topic modeling to understand how to increase user engagement~\cite{chen2013social}. However, a common problem with these recommendation algorithms is their propensity to create filter bubbles. 
While most research focuses on user interests, more abstract means of recommendation remains unexplored --such as building user networks based on temperament, conversation style, or other imaginative aspects -- remain unexplored.

While most research focuses on user interests, some studies are beginning to explore more abstract means of building online communities such as personality-driven recommendation systems \cite{dhelim2022survey, tkalvcivc2012personality}.
Currently a number of researchers are investigating ways to continue using topic-based recommendation algorithms that do not result in filter bubbles \cite{gao2023cirs, nguyen2014exploring, wang2022user}. However, there is ample room to challenge the foundational ideals which have led to their creation.

\subsection{Need for Interdisciplinary Work}
\label{ssec:gaps_2}

Given that online toxicity spans many disciplines, it is imperative that researchers across disciplines work together towards more holistic designs. Although the design of SMPs is primarily viewed as within the scope of traditional software engineering, sociologist, psychologists, political scientists, data scientists, among other need to be involved to address the wider problem of toxicity. This 
can lead to effective means to combat the many contributing factors of toxicity.


\subsection{Virtual versus Physical Interactions}
\label{ssec:gaps_3}

At the core of the problem of toxicity in SMPs is the notion that virtual spaces have evolved so rapidly that societal norms have been circumvented. As SMPs continue to quickly release new features hinging on the latest technologies, it becomes increasingly important to understand the critical components of physical communities in order to bring the positive pieces of these interactions to virtual spaces. 


%% file: sections/10_conclusion.tex
\section{Conclusion}
\label{sec:conclusion}

The current online environment does not give SMP providers enough incentive to shift their focus away from prioritizing the monetization of platforms.
Until SMPs are designed with intentionality around creating safe spaces for online communities, cyberbullying and other forms of toxicity will only tighten their hold. 
The contributions of this work are fourfold: systematization of knowledge of toxicity, formal definition of toxicity, identification of gaps in scientific community, and proposing research avenues for improving SMPs.

%% file: main.bbl

%% file: main.bbl
\begin{thebibliography}{106}


\ifx \showCODEN    \undefined \def \showCODEN     #1{\unskip}     \fi
\ifx \showDOI      \undefined \def \showDOI       #1{#1}\fi
\ifx \showISBNx    \undefined \def \showISBNx     #1{\unskip}     \fi
\ifx \showISBNxiii \undefined \def \showISBNxiii  #1{\unskip}     \fi
\ifx \showISSN     \undefined \def \showISSN      #1{\unskip}     \fi
\ifx \showLCCN     \undefined \def \showLCCN      #1{\unskip}     \fi
\ifx \shownote     \undefined \def \shownote      #1{#1}          \fi
\ifx \showarticletitle \undefined \def \showarticletitle #1{#1}   \fi
\ifx \showURL      \undefined \def \showURL       {\relax}        \fi
\providecommand\bibfield[2]{#2}
\providecommand\bibinfo[2]{#2}
\providecommand\natexlab[1]{#1}
\providecommand\showeprint[2][]{arXiv:#2}

\bibitem[Abro et~al\mbox{.}(2020)]%
        {abro2020automatic}
\bibfield{author}{\bibinfo{person}{Sindhu Abro}, \bibinfo{person}{Sarang Shaikh}, \bibinfo{person}{Zahid~Hussain Khand}, \bibinfo{person}{Ali Zafar}, \bibinfo{person}{Sajid Khan}, {and} \bibinfo{person}{Ghulam Mujtaba}.} \bibinfo{year}{2020}\natexlab{}.
\newblock \showarticletitle{Automatic hate speech detection using machine learning: A comparative study}.
\newblock \bibinfo{journal}{\emph{International Journal of Advanced Computer Science and Applications}} (\bibinfo{year}{2020}).
\newblock


\bibitem[Akuma et~al\mbox{.}(2022)]%
        {akuma2022comparing}
\bibfield{author}{\bibinfo{person}{Stephen Akuma}, \bibinfo{person}{Tyosar Lubem}, {and} \bibinfo{person}{Isaac~Terngu Adom}.} \bibinfo{year}{2022}\natexlab{}.
\newblock \showarticletitle{Comparing Bag of Words and TF-IDF with different models for hate speech detection from live tweets}.
\newblock \bibinfo{journal}{\emph{International Journal of Information Technology}} \bibinfo{volume}{14}, \bibinfo{number}{7} (\bibinfo{year}{2022}), \bibinfo{pages}{3629--3635}.
\newblock


\bibitem[Alba and Frenkel(2019)]%
        {alba_frenkel_2019}
\bibfield{author}{\bibinfo{person}{Davey Alba} {and} \bibinfo{person}{Sheera Frenkel}.} \bibinfo{year}{2019}\natexlab{}.
\newblock \bibinfo{title}{Russia tests new disinformation tactics in Africa to expand influence}.
\newblock
\newblock


\bibitem[Allcott et~al\mbox{.}(2019)]%
        {allcott2019trends}
\bibfield{author}{\bibinfo{person}{Hunt Allcott}, \bibinfo{person}{Matthew Gentzkow}, {and} \bibinfo{person}{Chuan Yu}.} \bibinfo{year}{2019}\natexlab{}.
\newblock \showarticletitle{Trends in the diffusion of misinformation on social media}.
\newblock \bibinfo{journal}{\emph{Research \& Politics}} (\bibinfo{year}{2019}).
\newblock


\bibitem[Almerekhi et~al\mbox{.}(2020)]%
        {almerekhi2020these}
\bibfield{author}{\bibinfo{person}{Hind Almerekhi}, \bibinfo{person}{Haewoon Kwak}, \bibinfo{person}{Joni Salminen}, {and} \bibinfo{person}{Bernard~J Jansen}.} \bibinfo{year}{2020}\natexlab{}.
\newblock \showarticletitle{Are these comments triggering? predicting triggers of toxicity in online discussions}. In \bibinfo{booktitle}{\emph{Proceedings of the web conference 2020}}. \bibinfo{pages}{3033--3040}.
\newblock


\bibitem[Aridor et~al\mbox{.}(2020)]%
        {aridor2020deconstructing}
\bibfield{author}{\bibinfo{person}{Guy Aridor}, \bibinfo{person}{Duarte Goncalves}, {and} \bibinfo{person}{Shan Sikdar}.} \bibinfo{year}{2020}\natexlab{}.
\newblock \showarticletitle{Deconstructing the filter bubble: User decision-making and recommender systems}. In \bibinfo{booktitle}{\emph{Proceedings of the 14th ACM conference on recommender systems}}. \bibinfo{pages}{82--91}.
\newblock


\bibitem[Au et~al\mbox{.}(2022)]%
        {au2022role}
\bibfield{author}{\bibinfo{person}{Cheuk~Hang Au}, \bibinfo{person}{Kevin~KW Ho}, {and} \bibinfo{person}{Dickson~KW Chiu}.} \bibinfo{year}{2022}\natexlab{}.
\newblock \showarticletitle{The role of online misinformation and fake news in ideological polarization: barriers, catalysts, and implications}.
\newblock \bibinfo{journal}{\emph{Information systems frontiers}} (\bibinfo{year}{2022}), \bibinfo{pages}{1--24}.
\newblock


\bibitem[Avalle et~al\mbox{.}(2024)]%
        {avalle2024persistent}
\bibfield{author}{\bibinfo{person}{Michele Avalle}, \bibinfo{person}{Niccol{\`o} Di~Marco}, \bibinfo{person}{Gabriele Etta}, \bibinfo{person}{Emanuele Sangiorgio}, \bibinfo{person}{Shayan Alipour}, \bibinfo{person}{Anita Bonetti}, \bibinfo{person}{Lorenzo Alvisi}, \bibinfo{person}{Antonio Scala}, \bibinfo{person}{Andrea Baronchelli}, \bibinfo{person}{Matteo Cinelli}, {et~al\mbox{.}}} \bibinfo{year}{2024}\natexlab{}.
\newblock \showarticletitle{Persistent interaction patterns across social media platforms and over time}.
\newblock \bibinfo{journal}{\emph{Nature}} \bibinfo{volume}{628}, \bibinfo{number}{8008} (\bibinfo{year}{2024}), \bibinfo{pages}{582--589}.
\newblock


\bibitem[Aydemir(2020)]%
        {aydemir2020social}
\bibfield{author}{\bibinfo{person}{Ilayda Aydemir}.} \bibinfo{year}{2020}\natexlab{}.
\newblock \showarticletitle{SOCIAL MEDIA AUTOMATION EFFECTS ON BRAND REPUTATION}.
\newblock  (\bibinfo{year}{2020}).
\newblock


\bibitem[Azzimonti and Fernandes(2022)]%
        {azzimonti2022social}
\bibfield{author}{\bibinfo{person}{Marina Azzimonti} {and} \bibinfo{person}{Marcos Fernandes}.} \bibinfo{year}{2022}\natexlab{}.
\newblock \showarticletitle{Social media networks, fake news, and polarization}.
\newblock \bibinfo{journal}{\emph{European Journal of Political Economy}} (\bibinfo{year}{2022}).
\newblock


\bibitem[Balkus and Yan(2022)]%
        {balkus2022improving}
\bibfield{author}{\bibinfo{person}{Salvador~V Balkus} {and} \bibinfo{person}{Donghui Yan}.} \bibinfo{year}{2022}\natexlab{}.
\newblock \showarticletitle{Improving short text classification with augmented data using GPT-3}.
\newblock \bibinfo{journal}{\emph{Natural Language Engineering}} (\bibinfo{year}{2022}), \bibinfo{pages}{1--30}.
\newblock


\bibitem[Banjo(2019)]%
        {washingtonpost}
\bibfield{author}{\bibinfo{person}{Shelly Banjo}.} \bibinfo{year}{2019}\natexlab{}.
\newblock \bibinfo{title}{Facebook, Twitter and the Digital Disinformation Mess}.
\newblock
\newblock


\bibitem[Barlett(2015)]%
        {barlett2015anonymously}
\bibfield{author}{\bibinfo{person}{Christopher~P Barlett}.} \bibinfo{year}{2015}\natexlab{}.
\newblock \showarticletitle{Anonymously hurting others online: The effect of anonymity on cyberbullying frequency.}
\newblock \bibinfo{journal}{\emph{Psychology of Popular Media Culture}} \bibinfo{volume}{4}, \bibinfo{number}{2} (\bibinfo{year}{2015}), \bibinfo{pages}{70}.
\newblock


\bibitem[Belanger(2021)]%
        {belanger2021comparison}
\bibfield{author}{\bibinfo{person}{Alyssa Belanger}.} \bibinfo{year}{2021}\natexlab{}.
\newblock \showarticletitle{The Comparison of Effectiveness of Social Media Marketing by Brands and Influencers for Organic Cosmetics}.
\newblock  (\bibinfo{year}{2021}).
\newblock


\bibitem[Bloch(2022)]%
        {bloch2022aversive}
\bibfield{author}{\bibinfo{person}{Stefano Bloch}.} \bibinfo{year}{2022}\natexlab{}.
\newblock \showarticletitle{Aversive racism and community-instigated policing: The spatial politics of Nextdoor}.
\newblock \bibinfo{journal}{\emph{Environment and Planning C: Politics and Space}} (\bibinfo{year}{2022}).
\newblock


\bibitem[Boyer(2021)]%
        {boyer_2021}
\bibfield{author}{\bibinfo{person}{Dustin Boyer}.} \bibinfo{year}{2021}\natexlab{}.
\newblock \bibinfo{title}{Streaming bots are ruining careers of indie musicians}.
\newblock
\newblock


\bibitem[Chen et~al\mbox{.}(2013)]%
        {chen2013social}
\bibfield{author}{\bibinfo{person}{Song Chen}, \bibinfo{person}{Samuel Owusu}, {and} \bibinfo{person}{Lina Zhou}.} \bibinfo{year}{2013}\natexlab{}.
\newblock \showarticletitle{Social network based recommendation systems: A short survey}. In \bibinfo{booktitle}{\emph{International conference on social computing}}. IEEE.
\newblock


\bibitem[Chitra and Musco(2020)]%
        {chitra2020analyzing}
\bibfield{author}{\bibinfo{person}{Uthsav Chitra} {and} \bibinfo{person}{Christopher Musco}.} \bibinfo{year}{2020}\natexlab{}.
\newblock \showarticletitle{Analyzing the impact of filter bubbles on social network polarization}. In \bibinfo{booktitle}{\emph{International Conference on Web Search and Data Mining}}. \bibinfo{pages}{115--123}.
\newblock


\bibitem[Chu et~al\mbox{.}(2024)]%
        {chu2024characterizing}
\bibfield{author}{\bibinfo{person}{Minh~Duc Chu}, \bibinfo{person}{Aryan Karnati}, \bibinfo{person}{Zihao He}, {and} \bibinfo{person}{Kristina Lerman}.} \bibinfo{year}{2024}\natexlab{}.
\newblock \showarticletitle{Characterizing Online Eating Disorder Communities with Large Language Models}.
\newblock \bibinfo{journal}{\emph{arXiv preprint arXiv:2401.09647}} (\bibinfo{year}{2024}).
\newblock


\bibitem[Cinelli et~al\mbox{.}(2021)]%
        {cinelli2021dynamics}
\bibfield{author}{\bibinfo{person}{Matteo Cinelli}, \bibinfo{person}{Andra{\v{z}} Pelicon}, \bibinfo{person}{Igor Mozeti{\v{c}}}, \bibinfo{person}{Walter Quattrociocchi}, \bibinfo{person}{Petra~Kralj Novak}, {and} \bibinfo{person}{Fabiana Zollo}.} \bibinfo{year}{2021}\natexlab{}.
\newblock \showarticletitle{Dynamics of online hate and misinformation}.
\newblock \bibinfo{journal}{\emph{Scientific reports}} (\bibinfo{year}{2021}).
\newblock


\bibitem[Cosma et~al\mbox{.}(2020)]%
        {cosma2020bullying}
\bibfield{author}{\bibinfo{person}{Alina Cosma}, \bibinfo{person}{Sophie~D Walsh}, \bibinfo{person}{Kayleigh~L Chester}, \bibinfo{person}{Mary Callaghan}, \bibinfo{person}{Michal Molcho}, \bibinfo{person}{Wendy Craig}, {and} \bibinfo{person}{William Pickett}.} \bibinfo{year}{2020}\natexlab{}.
\newblock \showarticletitle{Bullying victimization: Time trends and the overlap between traditional and cyberbullying across countries in Europe and North America}.
\newblock \bibinfo{journal}{\emph{International journal of public health}}  \bibinfo{volume}{65} (\bibinfo{year}{2020}), \bibinfo{pages}{75--85}.
\newblock


\bibitem[Croom(2011)]%
        {croom2011slurs}
\bibfield{author}{\bibinfo{person}{Adam~M Croom}.} \bibinfo{year}{2011}\natexlab{}.
\newblock \showarticletitle{Slurs}.
\newblock \bibinfo{journal}{\emph{Language Sciences}} (\bibinfo{year}{2011}).
\newblock


\bibitem[Dadvar et~al\mbox{.}(2013)]%
        {dadvar2013improving}
\bibfield{author}{\bibinfo{person}{Maral Dadvar}, \bibinfo{person}{Dolf Trieschnigg}, \bibinfo{person}{Roeland Ordelman}, {and} \bibinfo{person}{Franciska De~Jong}.} \bibinfo{year}{2013}\natexlab{}.
\newblock \showarticletitle{Improving cyberbullying detection with user context}. In \bibinfo{booktitle}{\emph{Advances in Information Retrieval: 35th European Conference on IR Research, ECIR 2013, Moscow, Russia, March 24-27, 2013. Proceedings 35}}. Springer, \bibinfo{pages}{693--696}.
\newblock


\bibitem[Dahlgren(2021)]%
        {dahlgren2021critical}
\bibfield{author}{\bibinfo{person}{Peter~M Dahlgren}.} \bibinfo{year}{2021}\natexlab{}.
\newblock \showarticletitle{A critical review of filter bubbles and a comparison with selective exposure}.
\newblock \bibinfo{journal}{\emph{Nordicom Review}} \bibinfo{volume}{42}, \bibinfo{number}{1} (\bibinfo{year}{2021}), \bibinfo{pages}{15--33}.
\newblock


\bibitem[Dhelim et~al\mbox{.}(2022)]%
        {dhelim2022survey}
\bibfield{author}{\bibinfo{person}{Sahraoui Dhelim}, \bibinfo{person}{Nyothiri Aung}, \bibinfo{person}{Mohammed~Amine Bouras}, \bibinfo{person}{Huansheng Ning}, {and} \bibinfo{person}{Erik Cambria}.} \bibinfo{year}{2022}\natexlab{}.
\newblock \showarticletitle{A survey on personality-aware recommendation systems}.
\newblock \bibinfo{journal}{\emph{Artificial Intelligence Review}} (\bibinfo{year}{2022}), \bibinfo{pages}{1--46}.
\newblock


\bibitem[Diaz(2021)]%
        {diaz_2021}
\bibfield{author}{\bibinfo{person}{Jaclyn Diaz}.} \bibinfo{year}{2021}\natexlab{}.
\newblock \bibinfo{title}{Want to send a mean tweet? Twitter's new feature wants you to think again}.
\newblock
\newblock


\bibitem[Ding et~al\mbox{.}(2020)]%
        {ding2020profiles}
\bibfield{author}{\bibinfo{person}{Yue Ding}, \bibinfo{person}{Dongping Li}, \bibinfo{person}{Xian Li}, \bibinfo{person}{Jiale Xiao}, \bibinfo{person}{Haiyan Zhang}, {and} \bibinfo{person}{Yanhui Wang}.} \bibinfo{year}{2020}\natexlab{}.
\newblock \showarticletitle{Profiles of adolescent traditional and cyber bullying and victimization: The role of demographic, individual, family, school, and peer factors}.
\newblock \bibinfo{journal}{\emph{Computers in Human Behavior}}  \bibinfo{volume}{111} (\bibinfo{year}{2020}), \bibinfo{pages}{106439}.
\newblock


\bibitem[Dutta et~al\mbox{.}(2021)]%
        {dutta2021analyzing}
\bibfield{author}{\bibinfo{person}{Upasana Dutta}, \bibinfo{person}{Rhett Hanscom}, \bibinfo{person}{Jason~Shuo Zhang}, \bibinfo{person}{Richard Han}, \bibinfo{person}{Tamara Lehman}, \bibinfo{person}{Qin Lv}, {and} \bibinfo{person}{Shivakant Mishra}.} \bibinfo{year}{2021}\natexlab{}.
\newblock \showarticletitle{Analyzing Twitter Users' Behavior Before and After Contact by the Russia's Internet Research Agency}.
\newblock \bibinfo{journal}{\emph{Human-Computer Interaction (HCI)}} (\bibinfo{year}{2021}).
\newblock


\bibitem[Eady et~al\mbox{.}(2023)]%
        {eady2023exposure}
\bibfield{author}{\bibinfo{person}{Gregory Eady}, \bibinfo{person}{Tom Paskhalis}, \bibinfo{person}{Jan Zilinsky}, \bibinfo{person}{Richard Bonneau}, \bibinfo{person}{Jonathan Nagler}, {and} \bibinfo{person}{Joshua~A Tucker}.} \bibinfo{year}{2023}\natexlab{}.
\newblock \showarticletitle{Exposure to the Russian Internet Research Agency foreign influence campaign on Twitter in the 2016 US election and its relationship to attitudes and voting behavior}.
\newblock \bibinfo{journal}{\emph{Nature Communications}} (\bibinfo{year}{2023}).
\newblock


\bibitem[Ehrett(2016)]%
        {ehrett2016judiciaries}
\bibfield{author}{\bibinfo{person}{John~S Ehrett}.} \bibinfo{year}{2016}\natexlab{}.
\newblock \showarticletitle{E-judiciaries: a model for community policing in cyberspace}.
\newblock \bibinfo{journal}{\emph{Information \& Communications Technology Law}} (\bibinfo{year}{2016}).
\newblock


\bibitem[Etta et~al\mbox{.}(2024)]%
        {etta2024topology}
\bibfield{author}{\bibinfo{person}{Gabriele Etta}, \bibinfo{person}{Matteo Cinelli}, \bibinfo{person}{Niccolo Di~Marco}, \bibinfo{person}{Michele Avalle}, \bibinfo{person}{Alessandro Panconesi}, {and} \bibinfo{person}{Walter Quattrociocchi}.} \bibinfo{year}{2024}\natexlab{}.
\newblock \showarticletitle{A Topology-Based Approach for Predicting Toxic Outcomes on Twitter and YouTube}.
\newblock \bibinfo{journal}{\emph{IEEE Transactions on Network Science and Engineering}} (\bibinfo{year}{2024}).
\newblock


\bibitem[Facebook(2023)]%
        {meta}
\bibfield{author}{\bibinfo{person}{Facebook}.} \bibinfo{year}{2023}\natexlab{}.
\newblock \bibinfo{title}{How to handle harassment and bullying on Facebook and Instagram}.
\newblock \bibinfo{howpublished}{\url{https://www.facebook.com/business/learn/lessons/handle-harassment-bullying}}.
\newblock


\bibitem[Fan et~al\mbox{.}(2021)]%
        {fan2021social}
\bibfield{author}{\bibinfo{person}{Hong Fan}, \bibinfo{person}{Wu Du}, \bibinfo{person}{Abdelghani Dahou}, \bibinfo{person}{Ahmed~A Ewees}, \bibinfo{person}{Dalia Yousri}, \bibinfo{person}{Mohamed~Abd Elaziz}, \bibinfo{person}{Ammar~H Elsheikh}, \bibinfo{person}{Laith Abualigah}, {and} \bibinfo{person}{Mohammed~AA Al-qaness}.} \bibinfo{year}{2021}\natexlab{}.
\newblock \showarticletitle{Social media toxicity classification using deep learning: real-world application UK brexit}.
\newblock \bibinfo{journal}{\emph{Electronics}} \bibinfo{volume}{10}, \bibinfo{number}{11} (\bibinfo{year}{2021}), \bibinfo{pages}{1332}.
\newblock


\bibitem[Fiske(2018)]%
        {fiske2018controlling}
\bibfield{author}{\bibinfo{person}{Susan~T Fiske}.} \bibinfo{year}{2018}\natexlab{}.
\newblock \showarticletitle{Controlling other people: The impact of power on stereotyping}.
\newblock In \bibinfo{booktitle}{\emph{Social cognition}}. \bibinfo{publisher}{Routledge}, \bibinfo{pages}{101--115}.
\newblock


\bibitem[Folmer-Annevelink et~al\mbox{.}(2010)]%
        {folmer2010class}
\bibfield{author}{\bibinfo{person}{Elvira Folmer-Annevelink}, \bibinfo{person}{Simone Doolaard}, \bibinfo{person}{Mayra Mascare{\~n}o}, {and} \bibinfo{person}{Roel~J Bosker}.} \bibinfo{year}{2010}\natexlab{}.
\newblock \showarticletitle{Class size effects on the number and types of student-teacher interactions in primary classrooms}.
\newblock \bibinfo{journal}{\emph{The Journal of Classroom Interaction}} (\bibinfo{year}{2010}).
\newblock


\bibitem[Gao et~al\mbox{.}(2023)]%
        {gao2023cirs}
\bibfield{author}{\bibinfo{person}{Chongming Gao}, \bibinfo{person}{Shiqi Wang}, \bibinfo{person}{Shijun Li}, \bibinfo{person}{Jiawei Chen}, \bibinfo{person}{Xiangnan He}, \bibinfo{person}{Wenqiang Lei}, \bibinfo{person}{Biao Li}, \bibinfo{person}{Yuan Zhang}, {and} \bibinfo{person}{Peng Jiang}.} \bibinfo{year}{2023}\natexlab{}.
\newblock \showarticletitle{CIRS: Bursting filter bubbles by counterfactual interactive recommender system}.
\newblock \bibinfo{journal}{\emph{ACM Transactions on Information Systems}} \bibinfo{volume}{42}, \bibinfo{number}{1} (\bibinfo{year}{2023}), \bibinfo{pages}{1--27}.
\newblock


\bibitem[Garaigordobil(2015)]%
        {garaigordobil2015cyberbullying}
\bibfield{author}{\bibinfo{person}{Maite Garaigordobil}.} \bibinfo{year}{2015}\natexlab{}.
\newblock \showarticletitle{Cyberbullying in adolescents and youth in the Basque Country: prevalence of cybervictims, cyberaggressors, and cyberobservers}.
\newblock \bibinfo{journal}{\emph{Journal of Youth Studies}} (\bibinfo{year}{2015}).
\newblock


\bibitem[Goyal et~al\mbox{.}(2022)]%
        {goyal2022your}
\bibfield{author}{\bibinfo{person}{Nitesh Goyal}, \bibinfo{person}{Ian~D Kivlichan}, \bibinfo{person}{Rachel Rosen}, {and} \bibinfo{person}{Lucy Vasserman}.} \bibinfo{year}{2022}\natexlab{}.
\newblock \showarticletitle{Is your toxicity my toxicity? exploring the impact of rater identity on toxicity annotation}.
\newblock \bibinfo{journal}{\emph{Human-Computer Interaction}} (\bibinfo{year}{2022}).
\newblock


\bibitem[Grossetti et~al\mbox{.}(2020)]%
        {grossetti2020community}
\bibfield{author}{\bibinfo{person}{Quentin Grossetti}, \bibinfo{person}{C{\'e}dric~du Mouza}, {and} \bibinfo{person}{Nicolas Travers}.} \bibinfo{year}{2020}\natexlab{}.
\newblock \showarticletitle{Community-based recommendations on Twitter: avoiding the filter bubble}. In \bibinfo{booktitle}{\emph{International Conference on Web Information Systems Engineering}}. Springer.
\newblock


\bibitem[Hanbury({[n.\,d.]})]%
        {hanbury}
\bibfield{author}{\bibinfo{person}{Mary Hanbury}.} \bibinfo{year}{[n.\,d.]}\natexlab{}.
\newblock \bibinfo{title}{Facebook overhauls group privacy settings and gives admins tools to scan rule-breaking posts in its war on toxic content}.
\newblock
\newblock


\bibitem[He et~al\mbox{.}(2023)]%
        {he2023you}
\bibfield{author}{\bibinfo{person}{Xinlei He}, \bibinfo{person}{Savvas Zannettou}, \bibinfo{person}{Yun Shen}, {and} \bibinfo{person}{Yang Zhang}.} \bibinfo{year}{2023}\natexlab{}.
\newblock \showarticletitle{You Only Prompt Once: On the Capabilities of Prompt Learning on Large Language Models to Tackle Toxic Content}. In \bibinfo{booktitle}{\emph{Symposium on Security and Privacy (SP)}}. IEEE Computer Society.
\newblock


\bibitem[Hosseini et~al\mbox{.}(2017)]%
        {hosseini2017deceiving}
\bibfield{author}{\bibinfo{person}{Hossein Hosseini}, \bibinfo{person}{Sreeram Kannan}, \bibinfo{person}{Baosen Zhang}, {and} \bibinfo{person}{Radha Poovendran}.} \bibinfo{year}{2017}\natexlab{}.
\newblock \showarticletitle{Deceiving google's perspective api built for detecting toxic comments}.
\newblock \bibinfo{journal}{\emph{arXiv:1702.08138}} (\bibinfo{year}{2017}).
\newblock


\bibitem[Interian et~al\mbox{.}(2023)]%
        {interian2023network}
\bibfield{author}{\bibinfo{person}{Ruben Interian}, \bibinfo{person}{Rusl{\'a}n G.~Marzo}, \bibinfo{person}{Isela Mendoza}, {and} \bibinfo{person}{Celso~C Ribeiro}.} \bibinfo{year}{2023}\natexlab{}.
\newblock \showarticletitle{Network polarization, filter bubbles, and echo chambers: an annotated review of measures and reduction methods}.
\newblock \bibinfo{journal}{\emph{International Transactions in Operational Research}} (\bibinfo{year}{2023}).
\newblock


\bibitem[Jain et~al\mbox{.}(2018)]%
        {jain2018adversarial}
\bibfield{author}{\bibinfo{person}{Edwin Jain}, \bibinfo{person}{Stephan Brown}, \bibinfo{person}{Jeffery Chen}, \bibinfo{person}{Erin Neaton}, \bibinfo{person}{Mohammad Baidas}, \bibinfo{person}{Ziqian Dong}, \bibinfo{person}{Huanying Gu}, {and} \bibinfo{person}{Nabi~Sertac Artan}.} \bibinfo{year}{2018}\natexlab{}.
\newblock \showarticletitle{Adversarial Text Generation for Google's Perspective API}. In \bibinfo{booktitle}{\emph{International Conference on Computational Science and Computational Intelligence (CSCI)}}. IEEE.
\newblock


\bibitem[Jhaver et~al\mbox{.}(2023)]%
        {jhaver2023personalizing}
\bibfield{author}{\bibinfo{person}{Shagun Jhaver}, \bibinfo{person}{Alice~Qian Zhang}, \bibinfo{person}{Quan~Ze Chen}, \bibinfo{person}{Nikhila Natarajan}, \bibinfo{person}{Ruotong Wang}, {and} \bibinfo{person}{Amy~X Zhang}.} \bibinfo{year}{2023}\natexlab{}.
\newblock \showarticletitle{Personalizing content moderation on social media: User perspectives on moderation choices, interface design, and labor}.
\newblock \bibinfo{journal}{\emph{Proceedings of the ACM on Human-Computer Interaction}} \bibinfo{volume}{7}, \bibinfo{number}{CSCW2} (\bibinfo{year}{2023}), \bibinfo{pages}{1--33}.
\newblock


\bibitem[Joachims(1998)]%
        {joachims1998text}
\bibfield{author}{\bibinfo{person}{Thorsten Joachims}.} \bibinfo{year}{1998}\natexlab{}.
\newblock \showarticletitle{Text categorization with support vector machines: Learning with many relevant features}. In \bibinfo{booktitle}{\emph{European conference on machine learning}}. Springer.
\newblock


\bibitem[Keijzer and M{\"a}s(2022)]%
        {keijzer2022complex}
\bibfield{author}{\bibinfo{person}{Marijn~A Keijzer} {and} \bibinfo{person}{Michael M{\"a}s}.} \bibinfo{year}{2022}\natexlab{}.
\newblock \showarticletitle{The complex link between filter bubbles and opinion polarization}.
\newblock \bibinfo{journal}{\emph{Data Science}} \bibinfo{volume}{5}, \bibinfo{number}{2} (\bibinfo{year}{2022}), \bibinfo{pages}{139--166}.
\newblock


\bibitem[Kim et~al\mbox{.}(2021)]%
        {kim2021distorting}
\bibfield{author}{\bibinfo{person}{Jin~Woo Kim}, \bibinfo{person}{Andrew Guess}, \bibinfo{person}{Brendan Nyhan}, {and} \bibinfo{person}{Jason Reifler}.} \bibinfo{year}{2021}\natexlab{}.
\newblock \showarticletitle{The distorting prism of social media: How self-selection and exposure to incivility fuel online comment toxicity}.
\newblock \bibinfo{journal}{\emph{Journal of Communication}} (\bibinfo{year}{2021}).
\newblock


\bibitem[Klein and O’Brien(2018)]%
        {klein2018people}
\bibfield{author}{\bibinfo{person}{Nadav Klein} {and} \bibinfo{person}{Ed O’Brien}.} \bibinfo{year}{2018}\natexlab{}.
\newblock \showarticletitle{People use less information than they think to make up their minds}.
\newblock \bibinfo{journal}{\emph{National Academy of Sciences}} (\bibinfo{year}{2018}).
\newblock


\bibitem[Kubin and von Sikorski(2021)]%
        {kubin2021role}
\bibfield{author}{\bibinfo{person}{Emily Kubin} {and} \bibinfo{person}{Christian von Sikorski}.} \bibinfo{year}{2021}\natexlab{}.
\newblock \showarticletitle{The role of (social) media in political polarization: a systematic review}.
\newblock \bibinfo{journal}{\emph{Annals of the International Communication Association}} (\bibinfo{year}{2021}).
\newblock


\bibitem[Kumar et~al\mbox{.}(2021)]%
        {kumar2021designing}
\bibfield{author}{\bibinfo{person}{Deepak Kumar}, \bibinfo{person}{Patrick~Gage Kelley}, \bibinfo{person}{Sunny Consolvo}, \bibinfo{person}{Joshua Mason}, \bibinfo{person}{Elie Bursztein}, \bibinfo{person}{Zakir Durumeric}, \bibinfo{person}{Kurt Thomas}, {and} \bibinfo{person}{Michael Bailey}.} \bibinfo{year}{2021}\natexlab{}.
\newblock \showarticletitle{Designing toxic content classification for a diversity of perspectives}. In \bibinfo{booktitle}{\emph{Seventeenth Symposium on Usable Privacy and Security (SOUPS 2021)}}. \bibinfo{pages}{299--318}.
\newblock


\bibitem[Kurwa(2019)]%
        {kurwa2019building}
\bibfield{author}{\bibinfo{person}{Rahim Kurwa}.} \bibinfo{year}{2019}\natexlab{}.
\newblock \showarticletitle{Building the digitally gated community: The case of Nextdoor}.
\newblock \bibinfo{journal}{\emph{Surveillance \& Society}} (\bibinfo{year}{2019}).
\newblock


\bibitem[Lakshmanan(2022)]%
        {lakshmanan2022quest}
\bibfield{author}{\bibinfo{person}{Laks~VS Lakshmanan}.} \bibinfo{year}{2022}\natexlab{}.
\newblock \showarticletitle{On a quest for combating filter bubbles and misinformation}. In \bibinfo{booktitle}{\emph{International Conference on Management of Data}}.
\newblock


\bibitem[Li et~al\mbox{.}(2024)]%
        {li2024hot}
\bibfield{author}{\bibinfo{person}{Lingyao Li}, \bibinfo{person}{Lizhou Fan}, \bibinfo{person}{Shubham Atreja}, {and} \bibinfo{person}{Libby Hemphill}.} \bibinfo{year}{2024}\natexlab{}.
\newblock \showarticletitle{“HOT” ChatGPT: The promise of ChatGPT in detecting and discriminating hateful, offensive, and toxic comments on social media}.
\newblock \bibinfo{journal}{\emph{ACM Transactions on the Web}} (\bibinfo{year}{2024}).
\newblock


\bibitem[Madhyastha et~al\mbox{.}(2023)]%
        {Madhyastha_Founta_Specia_2023}
\bibfield{author}{\bibinfo{person}{Pranava Madhyastha}, \bibinfo{person}{Antigoni Founta}, {and} \bibinfo{person}{Lucia Specia}.} \bibinfo{year}{2023}\natexlab{}.
\newblock \showarticletitle{A study towards contextual understanding of toxicity in online conversations}.
\newblock \bibinfo{journal}{\emph{Natural Language Engineering}} \bibinfo{volume}{29}, \bibinfo{number}{6} (\bibinfo{year}{2023}), \bibinfo{pages}{1538–1560}.
\newblock
\urldef\tempurl%
\url{https://doi.org/10.1017/S1351324923000414}
\showDOI{\tempurl}


\bibitem[Mall et~al\mbox{.}(2020)]%
        {mall2020four}
\bibfield{author}{\bibinfo{person}{Raghvendra Mall}, \bibinfo{person}{Mridul Nagpal}, \bibinfo{person}{Joni Salminen}, \bibinfo{person}{Hind Almerekhi}, \bibinfo{person}{Soon-Gyo Jung}, {and} \bibinfo{person}{Bernard~J Jansen}.} \bibinfo{year}{2020}\natexlab{}.
\newblock \showarticletitle{Four types of toxic people: characterizing online users’ toxicity over time}. In \bibinfo{booktitle}{\emph{Nordic Conference on Human-Computer Interaction: Shaping Experiences, Shaping Society}}.
\newblock


\bibitem[Moore et~al\mbox{.}(2021)]%
        {moore2021deliberation}
\bibfield{author}{\bibinfo{person}{Alfred Moore}, \bibinfo{person}{Rolf Fredheim}, \bibinfo{person}{Dominik Wyss}, {and} \bibinfo{person}{Simon Beste}.} \bibinfo{year}{2021}\natexlab{}.
\newblock \showarticletitle{Deliberation and identity rules: The effect of anonymity, pseudonyms and real-name requirements on the cognitive complexity of online news comments}.
\newblock \bibinfo{journal}{\emph{Political Studies}} (\bibinfo{year}{2021}).
\newblock


\bibitem[Muneer and Fati(2020)]%
        {muneer2020comparative}
\bibfield{author}{\bibinfo{person}{Amgad Muneer} {and} \bibinfo{person}{Suliman~Mohamed Fati}.} \bibinfo{year}{2020}\natexlab{}.
\newblock \showarticletitle{A comparative analysis of machine learning techniques for cyberbullying detection on Twitter}.
\newblock \bibinfo{journal}{\emph{Future Internet}} (\bibinfo{year}{2020}).
\newblock


\bibitem[Nechushtai and Lewis(2019)]%
        {nechushtai2019kind}
\bibfield{author}{\bibinfo{person}{Efrat Nechushtai} {and} \bibinfo{person}{Seth~C Lewis}.} \bibinfo{year}{2019}\natexlab{}.
\newblock \showarticletitle{What kind of news gatekeepers do we want machines to be? Filter bubbles, fragmentation, and the normative dimensions of algorithmic recommendations}.
\newblock \bibinfo{journal}{\emph{Computers in human behavior}}  \bibinfo{volume}{90} (\bibinfo{year}{2019}), \bibinfo{pages}{298--307}.
\newblock


\bibitem[Nelimarkka et~al\mbox{.}(2018)]%
        {nelimarkka2018social}
\bibfield{author}{\bibinfo{person}{Matti Nelimarkka}, \bibinfo{person}{Salla-Maaria Laaksonen}, {and} \bibinfo{person}{Bryan Semaan}.} \bibinfo{year}{2018}\natexlab{}.
\newblock \showarticletitle{Social media is polarized, social media is polarized: towards a new design agenda for mitigating polarization}. In \bibinfo{booktitle}{\emph{Designing Interactive Systems Conference}}.
\newblock


\bibitem[Nguyen et~al\mbox{.}(2014)]%
        {nguyen2014exploring}
\bibfield{author}{\bibinfo{person}{Tien~T Nguyen}, \bibinfo{person}{Pik-Mai Hui}, \bibinfo{person}{F~Maxwell Harper}, \bibinfo{person}{Loren Terveen}, {and} \bibinfo{person}{Joseph~A Konstan}.} \bibinfo{year}{2014}\natexlab{}.
\newblock \showarticletitle{Exploring the filter bubble: the effect of using recommender systems on content diversity}. In \bibinfo{booktitle}{\emph{Proceedings of the 23rd international conference on World wide web}}. \bibinfo{pages}{677--686}.
\newblock


\bibitem[Pascual-Ferr{\'a} et~al\mbox{.}(2021)]%
        {pascual2021toxicity}
\bibfield{author}{\bibinfo{person}{Paola Pascual-Ferr{\'a}}, \bibinfo{person}{Neil Alperstein}, \bibinfo{person}{Daniel~J Barnett}, {and} \bibinfo{person}{Rajiv~N Rimal}.} \bibinfo{year}{2021}\natexlab{}.
\newblock \showarticletitle{Toxicity and verbal aggression on social media: Polarized discourse on wearing face masks during the COVID-19 pandemic}.
\newblock \bibinfo{journal}{\emph{Big Data \& Society}} \bibinfo{volume}{8}, \bibinfo{number}{1} (\bibinfo{year}{2021}), \bibinfo{pages}{20539517211023533}.
\newblock


\bibitem[Patchin and Hinduja(2011)]%
        {patchin2011traditional}
\bibfield{author}{\bibinfo{person}{Justin~W Patchin} {and} \bibinfo{person}{Sameer Hinduja}.} \bibinfo{year}{2011}\natexlab{}.
\newblock \showarticletitle{Traditional and nontraditional bullying among youth: A test of general strain theory}.
\newblock \bibinfo{journal}{\emph{Youth \& Society}} \bibinfo{volume}{43}, \bibinfo{number}{2} (\bibinfo{year}{2011}), \bibinfo{pages}{727--751}.
\newblock


\bibitem[Pavlopoulos et~al\mbox{.}(2021)]%
        {pavlopoulos2021semeval}
\bibfield{author}{\bibinfo{person}{John Pavlopoulos}, \bibinfo{person}{Jeffrey Sorensen}, \bibinfo{person}{L{\'e}o Laugier}, {and} \bibinfo{person}{Ion Androutsopoulos}.} \bibinfo{year}{2021}\natexlab{}.
\newblock \showarticletitle{SemEval-2021 task 5: Toxic spans detection}. In \bibinfo{booktitle}{\emph{international workshop on semantic evaluation}}.
\newblock


\bibitem[Payne(2017)]%
        {payne2017welcome}
\bibfield{author}{\bibinfo{person}{Will Payne}.} \bibinfo{year}{2017}\natexlab{}.
\newblock \showarticletitle{Welcome to the polygon: Contested digital neighborhoods and spatialized segregation on Nextdoor}.
\newblock \bibinfo{journal}{\emph{Computational Culture}} (\bibinfo{year}{2017}).
\newblock


\bibitem[Perrin(2015)]%
        {perrin2015social}
\bibfield{author}{\bibinfo{person}{Andrew Perrin}.} \bibinfo{year}{2015}\natexlab{}.
\newblock \showarticletitle{Social media usage}.
\newblock \bibinfo{journal}{\emph{Pew research center}} (\bibinfo{year}{2015}).
\newblock


\bibitem[{Perspective API}(2021a)]%
        {perspective_2}
\bibfield{author}{\bibinfo{person}{{Perspective API}}.} \bibinfo{year}{2021}\natexlab{a}.
\newblock \bibinfo{title}{Case studies}.
\newblock \bibinfo{howpublished}{\url{https://perspectiveapi.com/case-studies/}}.
\newblock


\bibitem[{Perspective API}(2021b)]%
        {perspective_1}
\bibfield{author}{\bibinfo{person}{{Perspective API}}.} \bibinfo{year}{2021}\natexlab{b}.
\newblock \bibinfo{title}{Using machine learning to reduce toxicity online}.
\newblock \bibinfo{howpublished}{\url{https://perspectiveapi.com/}}.
\newblock


\bibitem[Pete et~al\mbox{.}(2020)]%
        {pete2020social}
\bibfield{author}{\bibinfo{person}{Ildiko Pete}, \bibinfo{person}{Jack Hughes}, \bibinfo{person}{Yi~Ting Chua}, {and} \bibinfo{person}{Maria Bada}.} \bibinfo{year}{2020}\natexlab{}.
\newblock \showarticletitle{A social network analysis and comparison of six dark web forums}. In \bibinfo{booktitle}{\emph{European symposium on security and privacy workshops}}. IEEE.
\newblock


\bibitem[Pichel et~al\mbox{.}(2021)]%
        {pichel2021bullying}
\bibfield{author}{\bibinfo{person}{Rafael Pichel}, \bibinfo{person}{Mair{\'e}ad Foody}, \bibinfo{person}{James O’Higgins~Norman}, \bibinfo{person}{Sandra Feij{\'o}o}, \bibinfo{person}{Jes{\'u}s Varela}, {and} \bibinfo{person}{Antonio Rial}.} \bibinfo{year}{2021}\natexlab{}.
\newblock \showarticletitle{Bullying, cyberbullying and the overlap: What does age have to do with it?}
\newblock \bibinfo{journal}{\emph{Sustainability}} \bibinfo{volume}{13}, \bibinfo{number}{15} (\bibinfo{year}{2021}), \bibinfo{pages}{8527}.
\newblock


\bibitem[Pidikiti et~al\mbox{.}(2020)]%
        {pidikiti2020understanding}
\bibfield{author}{\bibinfo{person}{Srihaasa Pidikiti}, \bibinfo{person}{Jason~Shuo Zhang}, \bibinfo{person}{Richard Han}, \bibinfo{person}{Tamara Lehman}, \bibinfo{person}{Qin Lv}, {and} \bibinfo{person}{Shivakant Mishra}.} \bibinfo{year}{2020}\natexlab{}.
\newblock \showarticletitle{Understanding how readers determine the legitimacy of online news articles in the era of fake news}. In \bibinfo{booktitle}{\emph{International Conference on Advances in Social Networks Analysis and Mining (ASONAM)}}. IEEE.
\newblock


\bibitem[Pidikiti et~al\mbox{.}(2022)]%
        {pidikiti2022understanding}
\bibfield{author}{\bibinfo{person}{Srihaasa Pidikiti}, \bibinfo{person}{Jason~Shuo Zhang}, \bibinfo{person}{Richard Han}, \bibinfo{person}{Tamara~Silbergleit Lehman}, \bibinfo{person}{Qin Lv}, {and} \bibinfo{person}{Shivakant Mishra}.} \bibinfo{year}{2022}\natexlab{}.
\newblock \showarticletitle{Understanding How Readers Determine the Legitimacy of Online Medical News Articles in the Era of Fake News}.
\newblock In \bibinfo{booktitle}{\emph{Disease Control Through Social Network Surveillance}}. \bibinfo{publisher}{Springer}.
\newblock


\bibitem[Rasmussen et~al\mbox{.}(2024)]%
        {rasmussen2024super}
\bibfield{author}{\bibinfo{person}{Stig Hebbelstrup~Rye Rasmussen}, \bibinfo{person}{Alexander Bor}, \bibinfo{person}{Mathias Osmundsen}, {and} \bibinfo{person}{Michael~Bang Petersen}.} \bibinfo{year}{2024}\natexlab{}.
\newblock \showarticletitle{‘Super-unsupervised’classification for labelling text: online political hostility as an illustration}.
\newblock \bibinfo{journal}{\emph{British Journal of Political Science}} \bibinfo{volume}{54}, \bibinfo{number}{1} (\bibinfo{year}{2024}), \bibinfo{pages}{179--200}.
\newblock


\bibitem[Recuero(2024)]%
        {recuero2024platformization}
\bibfield{author}{\bibinfo{person}{Raquel Recuero}.} \bibinfo{year}{2024}\natexlab{}.
\newblock \showarticletitle{The Platformization of Violence: Toward a Concept of Discursive Toxicity on Social Media}.
\newblock \bibinfo{journal}{\emph{Social Media+ Society}} \bibinfo{volume}{10}, \bibinfo{number}{1} (\bibinfo{year}{2024}), \bibinfo{pages}{20563051231224264}.
\newblock


\bibitem[Reddit(2023)]%
        {reddit1}
\bibfield{author}{\bibinfo{person}{Reddit}.} \bibinfo{year}{2023}\natexlab{}.
\newblock \bibinfo{title}{Content policy}.
\newblock \bibinfo{howpublished}{\url{https://www.redditinc.com/policies/content-policy}}.
\newblock


\bibitem[Rhodes(2022)]%
        {rhodes2022filter}
\bibfield{author}{\bibinfo{person}{Samuel~C Rhodes}.} \bibinfo{year}{2022}\natexlab{}.
\newblock \showarticletitle{Filter bubbles, echo chambers, and fake news: how social media conditions individuals to be less critical of political misinformation}.
\newblock \bibinfo{journal}{\emph{Political Communication}} (\bibinfo{year}{2022}).
\newblock


\bibitem[Rupapara et~al\mbox{.}(2021)]%
        {rupapara2021impact}
\bibfield{author}{\bibinfo{person}{Vaibhav Rupapara}, \bibinfo{person}{Furqan Rustam}, \bibinfo{person}{Hina~Fatima Shahzad}, \bibinfo{person}{Arif Mehmood}, \bibinfo{person}{Imran Ashraf}, {and} \bibinfo{person}{Gyu~Sang Choi}.} \bibinfo{year}{2021}\natexlab{}.
\newblock \showarticletitle{Impact of SMOTE on imbalanced text features for toxic comments classification using RVVC model}.
\newblock \bibinfo{journal}{\emph{IEEE Access}}  \bibinfo{volume}{9} (\bibinfo{year}{2021}), \bibinfo{pages}{78621--78634}.
\newblock


\bibitem[Salminen et~al\mbox{.}(2020a)]%
        {salminen2020developing}
\bibfield{author}{\bibinfo{person}{Joni Salminen}, \bibinfo{person}{Maximilian Hopf}, \bibinfo{person}{Shammur~A Chowdhury}, \bibinfo{person}{Soon-gyo Jung}, \bibinfo{person}{Hind Almerekhi}, {and} \bibinfo{person}{Bernard~J Jansen}.} \bibinfo{year}{2020}\natexlab{a}.
\newblock \showarticletitle{Developing an online hate classifier for multiple social media platforms}.
\newblock \bibinfo{journal}{\emph{Human-centric Computing and Information Sciences}} (\bibinfo{year}{2020}).
\newblock


\bibitem[Salminen et~al\mbox{.}(2020b)]%
        {salminen2020topic}
\bibfield{author}{\bibinfo{person}{Joni Salminen}, \bibinfo{person}{Sercan Seng{\"u}n}, \bibinfo{person}{Juan Corporan}, \bibinfo{person}{Soon-gyo Jung}, {and} \bibinfo{person}{Bernard~J Jansen}.} \bibinfo{year}{2020}\natexlab{b}.
\newblock \showarticletitle{Topic-driven toxicity: Exploring the relationship between online toxicity and news topics}.
\newblock \bibinfo{journal}{\emph{PloS one}} \bibinfo{volume}{15}, \bibinfo{number}{2} (\bibinfo{year}{2020}), \bibinfo{pages}{e0228723}.
\newblock


\bibitem[Sap et~al\mbox{.}(2021)]%
        {sap2021annotators}
\bibfield{author}{\bibinfo{person}{Maarten Sap}, \bibinfo{person}{Swabha Swayamdipta}, \bibinfo{person}{Laura Vianna}, \bibinfo{person}{Xuhui Zhou}, \bibinfo{person}{Yejin Choi}, {and} \bibinfo{person}{Noah~A Smith}.} \bibinfo{year}{2021}\natexlab{}.
\newblock \showarticletitle{Annotators with attitudes: How annotator beliefs and identities bias toxic language detection}.
\newblock \bibinfo{journal}{\emph{arXiv preprint arXiv:2111.07997}} (\bibinfo{year}{2021}).
\newblock


\bibitem[Saveski et~al\mbox{.}(2021)]%
        {saveski2021structure}
\bibfield{author}{\bibinfo{person}{Martin Saveski}, \bibinfo{person}{Brandon Roy}, {and} \bibinfo{person}{Deb Roy}.} \bibinfo{year}{2021}\natexlab{}.
\newblock \showarticletitle{The structure of toxic conversations on Twitter}. In \bibinfo{booktitle}{\emph{Proceedings of the web conference 2021}}. \bibinfo{pages}{1086--1097}.
\newblock


\bibitem[Seering(2020)]%
        {seering2020reconsidering}
\bibfield{author}{\bibinfo{person}{Joseph Seering}.} \bibinfo{year}{2020}\natexlab{}.
\newblock \showarticletitle{Reconsidering community self-moderation: the role of research in supporting community-based models for online content moderation}.
\newblock \bibinfo{journal}{\emph{ACM on Human-Computer Interaction}}  \bibinfo{volume}{4} (\bibinfo{year}{2020}), \bibinfo{pages}{107}.
\newblock


\bibitem[Shao et~al\mbox{.}(2018)]%
        {shao2018spread}
\bibfield{author}{\bibinfo{person}{Chengcheng Shao}, \bibinfo{person}{Giovanni~Luca Ciampaglia}, \bibinfo{person}{Onur Varol}, \bibinfo{person}{Kai-Cheng Yang}, \bibinfo{person}{Alessandro Flammini}, {and} \bibinfo{person}{Filippo Menczer}.} \bibinfo{year}{2018}\natexlab{}.
\newblock \showarticletitle{The spread of low-credibility content by social bots}.
\newblock \bibinfo{journal}{\emph{Nature communications}} \bibinfo{volume}{9}, \bibinfo{number}{1} (\bibinfo{year}{2018}), \bibinfo{pages}{1--9}.
\newblock


\bibitem[Shcherbakova and Nikiforchuk(2022)]%
        {shcherbakova2022social}
\bibfield{author}{\bibinfo{person}{Olena Shcherbakova} {and} \bibinfo{person}{Svitlana Nikiforchuk}.} \bibinfo{year}{2022}\natexlab{}.
\newblock \showarticletitle{Social media and filter bubbles}.
\newblock \bibinfo{journal}{\emph{Scientific Journal of Polonia University}} \bibinfo{volume}{54}, \bibinfo{number}{5} (\bibinfo{year}{2022}), \bibinfo{pages}{81--88}.
\newblock


\bibitem[Sheth et~al\mbox{.}(2022)]%
        {sheth2022defining}
\bibfield{author}{\bibinfo{person}{Amit Sheth}, \bibinfo{person}{Valerie~L Shalin}, {and} \bibinfo{person}{Ugur Kursuncu}.} \bibinfo{year}{2022}\natexlab{}.
\newblock \showarticletitle{Defining and detecting toxicity on social media: context and knowledge are key}.
\newblock \bibinfo{journal}{\emph{Neurocomputing}} (\bibinfo{year}{2022}).
\newblock


\bibitem[Simchon et~al\mbox{.}(2022)]%
        {simchon2022troll}
\bibfield{author}{\bibinfo{person}{Almog Simchon}, \bibinfo{person}{William~J Brady}, {and} \bibinfo{person}{Jay~J Van~Bavel}.} \bibinfo{year}{2022}\natexlab{}.
\newblock \showarticletitle{Troll and divide: the language of online polarization}.
\newblock \bibinfo{journal}{\emph{PNAS nexus}} \bibinfo{volume}{1}, \bibinfo{number}{1} (\bibinfo{year}{2022}), \bibinfo{pages}{pgac019}.
\newblock


\bibitem[Spohr(2017)]%
        {spohr2017fake}
\bibfield{author}{\bibinfo{person}{Dominic Spohr}.} \bibinfo{year}{2017}\natexlab{}.
\newblock \showarticletitle{Fake news and ideological polarization: Filter bubbles and selective exposure on social media}.
\newblock \bibinfo{journal}{\emph{Business information review}} \bibinfo{volume}{34}, \bibinfo{number}{3} (\bibinfo{year}{2017}), \bibinfo{pages}{150--160}.
\newblock


\bibitem[Sun et~al\mbox{.}(2019)]%
        {sun2019fine}
\bibfield{author}{\bibinfo{person}{Chi Sun}, \bibinfo{person}{Xipeng Qiu}, \bibinfo{person}{Yige Xu}, {and} \bibinfo{person}{Xuanjing Huang}.} \bibinfo{year}{2019}\natexlab{}.
\newblock \showarticletitle{How to fine-tune bert for text classification?}. In \bibinfo{booktitle}{\emph{Chinese computational linguistics: 18th China national conference, CCL 2019, Kunming, China, October 18--20, 2019, proceedings 18}}. Springer, \bibinfo{pages}{194--206}.
\newblock


\bibitem[Tkal{\v{c}}i{\v{c}} and Chen(2012)]%
        {tkalvcivc2012personality}
\bibfield{author}{\bibinfo{person}{Marko Tkal{\v{c}}i{\v{c}}} {and} \bibinfo{person}{Li Chen}.} \bibinfo{year}{2012}\natexlab{}.
\newblock \showarticletitle{Personality and recommender systems}.
\newblock In \bibinfo{booktitle}{\emph{Recommender systems handbook}}. \bibinfo{publisher}{Springer}, \bibinfo{pages}{757--787}.
\newblock


\bibitem[Tomlein et~al\mbox{.}(2021)]%
        {tomlein2021audit}
\bibfield{author}{\bibinfo{person}{Matus Tomlein}, \bibinfo{person}{Branislav Pecher}, \bibinfo{person}{Jakub Simko}, \bibinfo{person}{Ivan Srba}, \bibinfo{person}{Robert Moro}, \bibinfo{person}{Elena Stefancova}, \bibinfo{person}{Michal Kompan}, \bibinfo{person}{Andrea Hrckova}, \bibinfo{person}{Juraj Podrouzek}, {and} \bibinfo{person}{Maria Bielikova}.} \bibinfo{year}{2021}\natexlab{}.
\newblock \showarticletitle{An audit of misinformation filter bubbles on YouTube: Bubble bursting and recent behavior changes}. In \bibinfo{booktitle}{\emph{Conference on Recommender Systems}}.
\newblock


\bibitem[Twitter(2023a)]%
        {twitter1}
\bibfield{author}{\bibinfo{person}{Twitter}.} \bibinfo{year}{2023}\natexlab{a}.
\newblock \bibinfo{title}{How twitter handles abusive behavior}.
\newblock \bibinfo{howpublished}{\url{https://help.twitter.com/en/rules-and-policies/abusive-behavior}}.
\newblock


\bibitem[Twitter(2023b)]%
        {twitter2}
\bibfield{author}{\bibinfo{person}{Twitter}.} \bibinfo{year}{2023}\natexlab{b}.
\newblock \bibinfo{title}{Our range of enforcement options for violations}.
\newblock \bibinfo{howpublished}{\url{https://help.twitter.com/en/rules-and-policies/enforcement-options}}.
\newblock


\bibitem[Vaidya et~al\mbox{.}(2024)]%
        {vaidya2024analysing}
\bibfield{author}{\bibinfo{person}{Aatman Vaidya}, \bibinfo{person}{Seema Nagar}, {and} \bibinfo{person}{Amit~A Nanavati}.} \bibinfo{year}{2024}\natexlab{}.
\newblock \showarticletitle{Analysing the Spread of Toxicity on Twitter}. In \bibinfo{booktitle}{\emph{Joint International Conference on Data Science \& Management of Data}}. \bibinfo{pages}{118--126}.
\newblock


\bibitem[Varanasi(2022)]%
        {varanasi}
\bibfield{author}{\bibinfo{person}{Lakshmi Varanasi}.} \bibinfo{year}{2022}\natexlab{}.
\newblock \bibinfo{title}{Twitter bots appear to be be in line with the company's estimate of below 5\%}.
\newblock
\newblock


\bibitem[Vasconcellos et~al\mbox{.}(2023)]%
        {vasconcellos2023analyzing}
\bibfield{author}{\bibinfo{person}{Paulo Henrique~Santos Vasconcellos}, \bibinfo{person}{Pedro Di{\'o}genes de~Almeida Lara}, {and} \bibinfo{person}{Humberto~Torres Marques-Neto}.} \bibinfo{year}{2023}\natexlab{}.
\newblock \showarticletitle{Analyzing polarization and toxicity on political debate in brazilian TikTok videos transcriptions}. In \bibinfo{booktitle}{\emph{Web Science Conference}}. \bibinfo{pages}{33--42}.
\newblock


\bibitem[Vogels(2021)]%
        {vogels2021state}
\bibfield{author}{\bibinfo{person}{Emily~A Vogels}.} \bibinfo{year}{2021}\natexlab{}.
\newblock \showarticletitle{The state of online harassment}.
\newblock \bibinfo{journal}{\emph{Pew Research Center}} (\bibinfo{year}{2021}).
\newblock


\bibitem[Vosoughi et~al\mbox{.}(2018)]%
        {vosoughi2018spread}
\bibfield{author}{\bibinfo{person}{Soroush Vosoughi}, \bibinfo{person}{Deb Roy}, {and} \bibinfo{person}{Sinan Aral}.} \bibinfo{year}{2018}\natexlab{}.
\newblock \showarticletitle{The spread of true and false news online}.
\newblock \bibinfo{journal}{\emph{science}} \bibinfo{volume}{359}, \bibinfo{number}{6380} (\bibinfo{year}{2018}), \bibinfo{pages}{1146--1151}.
\newblock


\bibitem[Vraga and Bode(2020)]%
        {vraga2020defining}
\bibfield{author}{\bibinfo{person}{Emily~K Vraga} {and} \bibinfo{person}{Leticia Bode}.} \bibinfo{year}{2020}\natexlab{}.
\newblock \showarticletitle{Defining misinformation and understanding its bounded nature: Using expertise and evidence for describing misinformation}.
\newblock \bibinfo{journal}{\emph{Political Communication}} \bibinfo{volume}{37}, \bibinfo{number}{1} (\bibinfo{year}{2020}), \bibinfo{pages}{136--144}.
\newblock


\bibitem[Wang et~al\mbox{.}(2021b)]%
        {wang2021review}
\bibfield{author}{\bibinfo{person}{Le Wang}, \bibinfo{person}{Meng Han}, \bibinfo{person}{Xiaojuan Li}, \bibinfo{person}{Ni Zhang}, {and} \bibinfo{person}{Haodong Cheng}.} \bibinfo{year}{2021}\natexlab{b}.
\newblock \showarticletitle{Review of classification methods on unbalanced data sets}.
\newblock \bibinfo{journal}{\emph{Ieee Access}}  \bibinfo{volume}{9} (\bibinfo{year}{2021}), \bibinfo{pages}{64606--64628}.
\newblock


\bibitem[Wang et~al\mbox{.}(2022)]%
        {wang2022user}
\bibfield{author}{\bibinfo{person}{Wenjie Wang}, \bibinfo{person}{Fuli Feng}, \bibinfo{person}{Liqiang Nie}, {and} \bibinfo{person}{Tat-Seng Chua}.} \bibinfo{year}{2022}\natexlab{}.
\newblock \showarticletitle{User-controllable recommendation against filter bubbles}. In \bibinfo{booktitle}{\emph{Proceedings of the 45th international ACM SIGIR conference on research and development in information retrieval}}. \bibinfo{pages}{1251--1261}.
\newblock


\bibitem[Wang et~al\mbox{.}(2021a)]%
        {wang2021analyzing}
\bibfield{author}{\bibinfo{person}{Yichen Wang}, \bibinfo{person}{Richard Han}, \bibinfo{person}{Tamara Lehman}, \bibinfo{person}{Qin Lv}, {and} \bibinfo{person}{Shivakant Mishra}.} \bibinfo{year}{2021}\natexlab{a}.
\newblock \showarticletitle{Analyzing behavioral changes of Twitter users after exposure to misinformation}. In \bibinfo{booktitle}{\emph{International Conference on Advances in Social Networks Analysis and Mining (ASONAM)}}.
\newblock


\bibitem[Wolfowicz et~al\mbox{.}(2023)]%
        {wolfowicz2023examining}
\bibfield{author}{\bibinfo{person}{Michael Wolfowicz}, \bibinfo{person}{David Weisburd}, {and} \bibinfo{person}{Badi Hasisi}.} \bibinfo{year}{2023}\natexlab{}.
\newblock \showarticletitle{Examining the interactive effects of the filter bubble and the echo chamber on radicalization}.
\newblock \bibinfo{journal}{\emph{Journal of Experimental Criminology}} \bibinfo{volume}{19}, \bibinfo{number}{1} (\bibinfo{year}{2023}), \bibinfo{pages}{119--141}.
\newblock


\bibitem[Woolley and Howard(2017)]%
        {woolley2017computational}
\bibfield{author}{\bibinfo{person}{Samuel~C Woolley} {and} \bibinfo{person}{Philip Howard}.} \bibinfo{year}{2017}\natexlab{}.
\newblock \showarticletitle{Computational propaganda worldwide: Executive summary}.
\newblock  (\bibinfo{year}{2017}).
\newblock


\bibitem[Yang et~al\mbox{.}(2022)]%
        {yang2022investigating}
\bibfield{author}{\bibinfo{person}{Tingting Yang}, \bibinfo{person}{Heng Luo}, {and} \bibinfo{person}{Di Sun}.} \bibinfo{year}{2022}\natexlab{}.
\newblock \showarticletitle{Investigating the combined effects of group size and group composition in online discussion}.
\newblock \bibinfo{journal}{\emph{Active Learning in Higher Education}} (\bibinfo{year}{2022}).
\newblock


\bibitem[Zhang and Kizilcec(2014)]%
        {zhang2014anonymity}
\bibfield{author}{\bibinfo{person}{Kaiping Zhang} {and} \bibinfo{person}{Ren{\'e}~F Kizilcec}.} \bibinfo{year}{2014}\natexlab{}.
\newblock \showarticletitle{Anonymity in social media: Effects of content controversiality and social endorsement on sharing behavior}. In \bibinfo{booktitle}{\emph{International AAAI Conference on Weblogs and Social Media}}.
\newblock


\bibitem[Zimmerman and Ybarra(2016)]%
        {zimmerman2016online}
\bibfield{author}{\bibinfo{person}{Adam~G Zimmerman} {and} \bibinfo{person}{Gabriel~J Ybarra}.} \bibinfo{year}{2016}\natexlab{}.
\newblock \showarticletitle{Online aggression: The influences of anonymity and social modeling.}
\newblock \bibinfo{journal}{\emph{Psychology of Popular Media Culture}} \bibinfo{volume}{5}, \bibinfo{number}{2} (\bibinfo{year}{2016}), \bibinfo{pages}{181}.
\newblock


\end{thebibliography}
